\documentclass{pasj01}

\usepackage{color}
\usepackage{natbib}

\Received{$\langle$reception date$\rangle$}
\Accepted{$\langle$acception date$\rangle$}
\Published{$\langle$publication date$\rangle$}

 \newcommand{\eco}{${\rm C^{18}O}$\,} 	
 \newcommand{\nnh}{N$_2$H$^{+}$\,} 	
 \newcommand{\Hii}{H{$\,${\sc ii}}~} 

\begin{document}

\title{Cluster formation in the W40 and Serpens South complex triggered by the expanding H{$\,${\sc ii}} region}

\author{Tomomi \textsc{Shimoikura}\altaffilmark{1}, Kazuhito \textsc{Dobashi}\altaffilmark{1},
Fumitaka \textsc{Nakamura}\altaffilmark{2,3,4}, Yoshito \textsc{Shimajiri}\altaffilmark{5}, and Koji \textsc{Sugitani}\altaffilmark{6}}
\altaffiltext{1}{Department of Astronomy and Earth Sciences, Tokyo Gakugei University, Koganei, Tokyo  184-8501, Japan} 
\altaffiltext{2}{National Astronomical Observatory of Japan, Mitaka, Tokyo 181-8588, Japan}
\altaffiltext{3}{Department of Astronomical Science, School of Physical Science, SOKENDAI (The Graduate University for Advanced Studies), Osawa, Mitaka, Tokyo 181-8588, Japan}
\altaffiltext{4}{Department of Astronomy, School of Science, University of Tokyo, Bunkyo, Tokyo, 113-0033, Japan}
\altaffiltext{5}{Laboratoire AIM, CEA/DSM-CNRS-Universit$\acute{\rm e}$ Paris Diderot, IRFU/Service d'Astrophysique, CEA Saclay, F-91191 Gif-sur-Yvette, France}
\altaffiltext{6}{Graduate School of Natural Sciences, Nagoya City University, Mizuho-ku, Nagoya 467-8501, Japan}
\email{ikura@u-gakugei.ac.jp}


\maketitle


\begin{abstract}
We present results of the mapping observations covering a large area of 1 square degree around
W40 and Serpens South carried out in the $^{12}$CO ($J=1-0$), $^{13}$CO ($J=1-0$), C$^{18}$O ($J=1-0$), CCS ($J_{N}$=8$_{7}-7_{6}$), and N$_2$H$^{+}$ ($J=1-0$)
emission lines with the 45 m Nobeyama Radio Telescope.
W40 is a blistered \Hii region, and Serpens South is an infrared dark cloud accompanied by a young cluster. 
The relationship between these two regions which are separated by $\sim20\arcmin$ on the sky
has not been clear so far.
We found that the C$^{18}$O emission is distributed smoothly throughout the W40 and Serpens South regions,
and it seems that the two regions are physically connected.
We divided the C$^{18}$O emission into four groups in terms of the spatial distributions around the \Hii region
which we call 5, 6, 7, and 8 km s$^{-1}$ components according to their typical LSR velocities,
and propose a three-dimensional model of the W40 and Serpens South complex.
We found two elliptical structures in position-velocity diagrams, which can be explained as a part of two expanding shells.
One of the shells is the small inner shell just around the \Hii region, and the other is the large outer shell corresponding to the boundary of the \Hii region.
Dense gas associated with the young cluster of Serpens South is likely to be located at the surface of the outer shell, indicating that the natal clump of the young cluster is interacting with the outer shell being compressed by the expansion of the shell. 
We suggest that the expansion of the shell induced the formation of the young cluster.
\end{abstract}


\section{ Introduction }
OB stars give significant influence on the physical and chemical environments of the natal clumps, and trigger formation of the new generation of stars \citep[e.g.,][]{Elmegreen}. Investigation of star formation around the \Hii regions is therefore essential to understand how star formation occurs and evolves by the influence of \Hii regions.

W40 and Serpens South, a part of the Aquila cloud complex, are deeply embedded in dust and gas.
Several infrared (IR) observations have revealed the global dust distributions of the two regions.
In figure \ref{fig:spitzer}, we show an overview of the two regions in a three-color composite image of the 8.0 \micron$\,$(red), 5.8 \micron$\,$(green) and 4.5 \micron$\,$(blue) emission from the $\it{Spitzer}$ $\it{Space}$ $\it{Telescope}$ (hereafter, $\it{Spitzer}$).
Recent VLBA parallax measurements by \cite{Ortiz2017} indicate a distance of $436\pm9$ pc for both W40 and Serpens South.

W40 is a blister-type \Hii region \cite[e.g.,][]{Vallee} excited by a massive OB star cluster and is characterized by the `hourglass-shaped structure' \citep{Rodney} 
which was revealed by  
infrared observations using the $\it{Midcourse}$ $\it{Space}$ $\it{Experiment}$ and $\it{Spitzer}$.
The hourglass-shaped structure consists of two interconnected cavities having an extent of $17\arcmin\times30\arcmin$ oriented with one lobe to the southeast and the other to the northwest \citep{Rodney}.
\cite{Shuping2012} conclude that the O9.5 star named IRS 1A South is the dominant source of Lyman continuum luminosity needed to power the \Hii region, 
and is the likely source of the stellar wind that has made the hourglass-shaped structure.
The dynamical age of the \Hii region has been estimated to be $0.19 - 0.78$ Myr \citep{Mallick}.
There are dense molecular clumps in and around the \Hii region \citep[e.g.,][]{Dobashi2005, Dobashi2011,Shimoikura2015,Rumble2016,Shimajiri2017}.
In our previous study with the $^{12}$CO ($J=3-2$) and HCO$^{+}$ ($J=4-3$) emission lines \citep{Shimoikura2015}, we presented an evidence that the velocity field of the region shows a great complexity consisting of at least four distinct velocity components, and two of the components at 5 km s$^{-1}$ and 10 km s$^{-1}$ are likely to be tracing dense gas interacting with the expanding shell around the \Hii region.
Hundreds of young stellar objects (YSOs) are associated with the region, indicating active on-going star formation \cite[e.g.,][]{Kuhn,Bontemps2010,Maury2011,Mallick}.
\cite{Kuhn} suggested that the YSOs are young with an age of $\leq1$ Myr.
These results suggest that the YSOs may be the second generation stars born by the influence of the expanding shell
of W40.
However, the previous studies were limited to a small area just around the \Hii region, not covering the entire W40 region. 
It is therefore not yet evident how far the interstellar medium around W40 is affected by the ionization front.

Toward the $\sim20^\prime$ west to the W40 region, there is another star-forming region Serpens South.
This region can be recognized as a filamentary obscuring structure in the $\it{Spitzer}$ image (see figure \ref{fig:spitzer}).
A young cluster has been found around the center of the structure \citep{Gutermuth2008}.
The cluster consists of a large fraction of protostars, 
some of which blow out the powerful collimated outflows \citep{Nakamura2011},
indicating that they formed within 0.5 Myr in the past \citep{Gutermuth2008}.
The filamentary structure was found to consist of multiple smaller filaments identified by gas and dust observations \citep[e.g.,][]{Kirk2013,Nakamura2014,Fernandez2014,Konyves2015}. 
These filaments show velocity gradients along their elongation \cite[e.g.,][]{Kirk2013,Nakamura2014,Fernandez2014}.
\cite{Kirk2013} suggested that the velocity gradient in one of the filaments is due to gas accretion toward the central young cluster.
On the other hand, \cite{Fernandez2014} pointed out that the velocity gradients along the filaments can be interpreted as a projection of large-scale turbulence, as they also found an existence of velocity gradients perpendicular to the main axes of the filaments. 
On a larger scale, results of low resolution molecular observations ($\sim 3\arcmin$) covering both W40 and Serpens South by \cite{Nakamura2017} infer that the two regions might be physically connected.
In that case, we wonder how much the Serpens South region can be influenced by the W40 \Hii region.
Though each region has been widely studied individually, their relationship has remained unclear to date.

The IR observations as shown in figure \ref{fig:spitzer} can reveal only a part of the relationship of W40 and Serpens South, 
and they provide information neither on the internal structure nor the velocity field.
To better illustrate the larger-scale view of the surroundings of W40 and Serpens South and to better understand the relationship between the two regions, large-scale mapping observations in molecular emission lines with a high angular resolution is needed.

In this work, we mapped the whole region shown in figure \ref{fig:spitzer} in some molecular lines
using the 45 m telescope at Nobeyama Radio Observatory (NRO).
This study is based on `the Star Formation Legacy project' which is a large-scale survey project of molecular gas in star-forming regions. 
The outline of the project will be described in a forthcoming paper (Nakamura et al. 2018, in preparation).
Results of wide-field polarimetric observations toward Serpens South is given in Kusune et al. (2018, in preparation; 
see also Sugitani et al. 2011).
Results of molecular line observations toward other star-forming regions will be presented
in separate articles (Dobashi et al. 2018, in preparation: Nakamura et al. 2018, in preparation: Shimoikura et al. 2018b, in preparation: Tanabe et al. 2018, in preparation ).

The present study intends to understand the complex spatial and velocity structures
of both the W40 and Serpens South regions by analyzing the molecular data.
We adopt a distance of 436 pc to the observed area \citep{Ortiz2017} in this paper.

We present the observational procedures in Section 2.
In Section 3, we show the global distributions of the molecular gas.
We also describe results of the analyses of the temperature distributions
and velocity fields.
In addition, we investigate the morphology and kinematics of the observed regions.
Our conclusions are summarized in Section 4.


\section{ Observations }
 
\subsection{ Observations with the NRO 45 m telescope }
Molecular observations were performed by using the NRO 45 m telescope. 
The observations were carried out for $\sim150$ hours in total in the period from 2015 April to 2017 March.
The $^{12}$CO (115.271204 GHz), 
$^{13}$CO (110.201354 GHz), C$^{18}$O (109.782176 GHz), CCS (93.870098GHz), and N$_2$H$^{+}$ (93.1737637 GHz) emission lines were observed simultaneously using the on-the-fly (OTF) mode.
The half-power beam width of the telescope at 110 GHz is $\sim15\arcsec$, corresponding to $\sim6500$ au at a distance of 436 pc.
We mapped an area of $\sim1$ square degree around W40 and Serpens South in these lines. 

We used the multi-beam receiver ``FOREST" \citep[FOur beam REceiver System on the 45-m Telescope,][]{Minamidani} as the frontend to obtain the large maps.
The typical system temperatures were 170 K at 110 GHz.
The backend used was the digital spectrometers SAM45 having a bandwidth of 31.25 MHz and a frequency resolution of 15.26 kHz.
The frequency resolution corresponds to a velocity resolution of $\sim0.04$ km s$^{-1}$ at 110 GHz.
For intensity calibration, the chopper-wheel method \citep{Kutner} was used.
The telescope pointing was checked by observing the SiO maser source V1111-Oph every 1.5 hour, 
which was found to be accurate within $\sim5\arcsec$.
We also observed a small region ($\sim1\farcm5 \times 1\farcm5$) in Serpens South 
every time we tuned the receiver to check the intensity calibration,  
and we found that the intensity fluctuations for all of the lines are less than $10 \%$.

The spectra were reduced by fitting the baseline with linear functions, 
and the data were converted to 3-dimensional data with a
spheroidal convolution at an angular grid of $7\farcs5$ and a velocity resolution of 0.1 km s$^{-1}$. 
The velocity resolution was smoothed to this value
to achieve a high signal to noise ratio.
We used the NOSTAR software package \citep{Sawada} for the reduction.
We finally obtained the spectral data with an effective angular resolution of $\sim22\arcsec$ at 110 GHz. 

Following the calibration formulae available at the NRO website\footnote{https://www.nro.nao.ac.jp/~nro45mrt/html/prop/eff/eff-intp.html},
we estimated the main beam efficiencies of the 45 m telescope for the
$^{12}$CO, $^{13}$CO, C$^{18}$O, CCS, and \nnh lines
to be 0.416, 0.435, 0.437, 0.497, and 0.500, respectively,
and used them
to convert the obtained $T_{\rm a}^*$ scale data  to $T_{\rm mb}$ scale data.
One sigma noise level of the reduced data is $\Delta T_{\rm mb}=0.8-0.3$ K for the
$\sim 0.1$ km s$^{-1}$
velocity resolution. The observed molecular lines and the resulting noise levels
are summarized in table \ref{tab:line}.

\subsection{ Archival data }
We used $\it{Herschel}$ archive data from $\it{Herschel}$ Gould Belt survey (HGBS), i.e.,
the 70, 250, 500 $\micron$ maps,
dust temperature map, and column density map of the observed region
 \citep{Andre2010, Konyves2015}.
We also used $\it{Spitzer}$ data obtained from the public $\it{Spitzer}$ archive\footnote{http://archive. spitzer.caltech.edu/}.
These infrared data are used to compare with the molecular lines data.

\section{ Results and Discussion }

\subsection{ The molecular distributions }

In figure \ref{fig:spe}, we show the $^{12}$CO, $^{13}$CO, and C$^{18}$O spectra averaged over
the observed region. 
The C$^{18}$O spectrum has a single peak at $V_{\rm LSR}=7.3$ km s$^{-1}$. 
There is a dip in the $^{12}$CO and $^{13}$CO spectra at this velocity,
which should be due to absorption by colder gas in the foreground.

Figure \ref{fig:iimap} shows the integrated intensity maps of the C$^{18}$O, N$_2$H$^{+}$, and CCS emission lines for the W40 and Serpens South regions.
As seen in figure \ref{fig:iimap}a, the C$^{18}$O distributions
around the densest parts of the W40 and Serpens South regions are similar to the structures found in our previous studies \cite[e.g.,][]{Shimoikura2015, Nakamura2014}. The figure clearly shows that the C$^{18}$O emission is
distributed over the entire observed area continuously. In addition, as we will show later,
radial velocities of the emission line change smoothly over the observed
area (e.g., see figure \ref{fig:PV1}).
We therefore conclude that the W40 and Serpens South
regions are physically connected. 

In figure \ref{fig:iimap}b, we also show the C$^{18}$O intensity map overlaid on the composite image of $\it{Herschel}$ $500$ $\micron$ (red), $70$ $\micron$ (green), and $\it{Spitzer}$ $8$ $\micron$ (blue).
The C$^{18}$O emission is well correlated spatially with the dust emission revealed by $\it{Herschel}$ showing several dust filaments.
We also show a composite image around W40 made from $\it{Spitzer}$ $8$ $\micron$, 2MASS $K$s, and $H$ bands in figure \ref{fig:dark}. 
We can see some dark lanes absorbed by dust, which can be seen in the dust emission in the $\it{Herschel}$ image.
These dust filaments seem to be connected with Serpens South across the \Hii region,
and are associated with the C$^{18}$O emission.
One of the filaments is located at the waist of the hourglass structure, and it is known that many YSOs are associated with this dark lane \citep[e.g.,][]{Kuhn,Maury2011,Mallick} as seen in figure \ref{fig:YSO} which shows the distributions of young cores around W40 and Serpens South cataloged by \cite{Andre2010} and \cite{Konyves2015}.

In figure \ref{fig:iimap}b, we also found that the \Hii region is sharply outlined by a cavity of the C$^{18}$O emission:
There is less C$^{18}$O emission in the southeast side of the W40 region as well as in the region between W40 and Serpens South, and such regions coincide well with the extent of the \Hii
region seen as the bright blue region in the $\it{Spitzer}$ image.

In figures \ref{fig:iimap}c and \ref{fig:iimap}d, elongated filamentary structures are detected with the CCS and \nnh emission lines.
The CCS and \nnh emission lines are dominant toward a part corresponding to the main body of Serpens South.
The morphology of the \nnh emission around the main body well matches with that of the dust emission
detected by $Herschel$ (shown in red in figure \ref{fig:iimap}b), which is consistent with the previous studies \citep{Kirk2013, Nakamura2014}.
We also found that the region showing the \nnh emission around the OB cluster is compact,
suggesting that the \nnh emission is enhanced around the OB cluster due to the high temperatures.
The CCS emission is not detected in the W40 region above the $3 \sigma$ noise level, as we
already found in our previous study with CCS ($J_N=4_3-3_2$) \citep{Shimoikura2015}. 
In Serpens South, we found that the CCS intensity peak does not coincide with the \nnh peak
corresponding to the position of the young cluster.
The CCS emission is the strongest at $\sim8\arcmin$ north to the \nnh intensity peak, and from there,
it extends to the northwest in the map (figure \ref{fig:iimap}c).
Observations of HC${_7}$N by \cite{Friesen} also show that the strong molecular
emission extends to the north of the young cluster, showing similar distributions seen in our CCS map.

In earlier studies, the CCS and \nnh emission lines were surveyed only in a limited region around the OB cluster in W40 \cite[e.g.,][]{Pirogov2013, Shimoikura2015} or around the main body of the Serpens South \cite[e.g.,][]{Kirk2013, Nakamura2014}. 
Our maps reveal the overall distributions of these molecular lines in W40 and Serpens South, and
we found that they are detected at some positions along the dust filaments as well as in dust condensations in the \Hii region, not only in the main body of the Serpens South.

From the above results, we found that the ionization front clearly delineates the boundary between the ionized region and the molecular gas. 
The morphology of the C$^{18}$O distribution shows that 
W40 and Serpens South are molecular clouds in the same system, and that
they are shaped by the expansion of the \Hii region.

In addition, we found an extended C$^{18}$O emitting region which may be a reflection nebula in the west of Serpens South.
The distribution is consistent with one of the $\it{Herschel}$ dust filaments, and only a part of it is detected in \nnh and CCS on the north side of the filament.


\subsection{The temperature distributions}\label{sec:temperature}

In figure \ref{fig:iimap}, we found that there is an anti-correlation between the C$^{18}$O emission and the $\it{Spitzer}$ image, and the molecular gas seems to be influenced by the \Hii region.
To investigate the influence of the \Hii region, we first estimated the excitation temperature $T_{\rm ex}$ for each observed molecular line throughout the observed region.
For the estimation, 
we assumed the Local Thermodynamic Equilibrium (LTE) and that the observed emission lines are optically thin.

Under the LTE condition, $T_{\rm mb}$ is expressed by the following radiative transfer equation,
\begin{equation}
\label{eq:radiative}
T_ {\rm mb}=[J(T_ {\rm ex})-J(T_{\rm bg})][1-{\rm exp}(-\tau(v))]\ 
\end{equation}
where  $J(T)=T_{0} /(e^{T_{0} /T}-1)$, $T_{0}=h\nu /k$, and $T_{\rm bg}=2.73$ K.
Here, $\nu$ is the rest frequency of the emission line, $k$ is the Boltzmann constant, $h$ is the Planck constant,
and $\tau(v)$ is the optical depth as a function of velocity.

We estimated $T_{\rm ex}$ of C$^{18}$O that we denoted as $T^{\rm{co}}_{\rm ex}$
from the $T_{\rm mb}$ data of the optically thick $^{12}$CO line ($\tau\gg1$) by measuring the maximum value in the $^{12}$CO spectra. 
Line parameters of the molecules were taken from Splatalogue\footnote{www.splatalogue.net}.
The $T^{\rm{co}}_{\rm ex}$ map obtained by using equation  (\ref{eq:radiative}) is shown in figure \ref{fig:temp}a.
In figure \ref{fig:temp}b, we also show the dust temperature map $T_{\rm dust}$
for comparison,
which is obtained by the $\it{Herschel}$ observations \citep{Andre2010,Konyves2015}.

For the \nnh molecule, we estimated the line parameters including the excitation temperature $T^{\rm{N{_2}H^{+}}}_{\rm ex}$ as follows:
The N$_2$H$^{+}$ ($J=1-0$) line has seven hyperfine components \citep[e.g.,][]{Caselli1995}. 
Because all of the components have the same $T^{\rm{N{_2}H^{+}}}_{\rm ex}$ and line width ${\Delta V}$, we performed a simultaneous fit with seven Gaussian components to the observed N$_2$H$^{+}$ ($J=1-0$) spectrum at each position using equation (\ref{eq:radiative}).
Here, $\tau(v)$ for the N$_2$H$^{+}$  line can be expressed as
\begin{equation}
\label{eq:nnh}
{\tau}(v)={\tau_{\rm tot}}\sum_{i=1}^{7} C_{i} {\rm exp} \frac{{-(v-V_{0}+v_{i})^{2}}}{{2\,\sigma_{i}}^{2}}
\end{equation}
where $C_{i}$ is the normalized relative intensity (i.e., $\sum_{i=1}^{7} C_{i}=1$) and
$v_{i}$ is the artificial velocity difference due to the slight differences of the rest frequencies.
The parameters for the estimations are taken from \cite{Tine2000} and Splatalogue.
 $\sigma_{i}$ is the velocity dispersion that can be expressed as
 ${\Delta V}/\sqrt{8{\rm ln}2}$ where $\Delta V$ is the
velocity dispersion measured at full width at half maximum (FWHM).
We selected positions where the N$_2$H$^{+}$ emission is detected with $\ge10$ $\sigma$,
and performed the fitting to derive 
the total optical depth of the seven hyperfine components $\tau_{\rm tot}$,  
the peak LSR velocity $V_{0}$, ${\Delta V}$, $T_{\rm mb}$,
and $T^{\rm{N{_2}H^{+}}}_{\rm ex}$ of each position. 
We show the fitted values of $T^{\rm{N{_2}H^{+}}}_{\rm ex}$ across the observed region in figure \ref{fig:temp}c.
In figure \ref{fig:spectra}, we also show an example of the observed and fitted \nnh spectra
together with the \eco and CCS spectra measured at some positions. 
Here, the derived parameters of \nnh at the position shown in figure \ref{fig:spectra}c
are $V_{0}=7.66\pm0.01$ km s$^{-1}$, $\tau_{\rm tot}=3.18\pm0.06$, 
${\Delta V}=1.72\pm0.03$ km s$^{-1}$, and $T^{\rm{N{_2}H^{+}}}_{\rm ex}=10.16\pm0.40$ K.
Similarly, the parameters at the position shown in figure \ref{fig:spectra}e are $V_{0}=4.65\pm0.01$ km s$^{-1}$, $\tau_{\rm tot}=4.24\pm0.11$, 
${\Delta V}=0.70\pm0.01$ km s$^{-1}$, and $T^{\rm{N{_2}H^{+}}}_{\rm ex}=8.47\pm0.70$ K.

In figure \ref{fig:temp}a, the $T^{\rm{co}}_{\rm ex}$ reaches a value of $56$ K in the vicinity of the OB cluster, 
whereas $T^{\rm{co}}_{\rm ex}$ around the Serpens region stays around $\sim10$ K 
and exceeds 15 K only around the young cluster.
We note that the derived values of $T^{\rm{co}}_{\rm ex}$ are the lower limits since the $^{12}$CO line often shows heavy
self-absorption around $\sim7$ km s$^{-1}$ as seen in figure \ref{fig:spe}. 
The $T_{\rm dust}$ map clearly shows that the distribution is similar to the shape of the \Hii region traced by $\it{Spitzer}$,
showing decreasing temperature with increasing distance from the ionized gas. 
Here, \nnh will be destroyed by CO evaporated from grain surfaces at 25 K \citep[e.g.,][]{Lee_2004}.
We found that $T_{\rm dust}$ around the OB cluster is higher than 25 K. 
As shown in figure \ref{fig:iimap}d, the \nnh emission around the OB cluster is compact,
and we thus suggest that the \nnh molecule is largely destroyed by CO evaporated from dust.
In the Serpens South region, there is a tendency that the temperature becomes higher around the young cluster in figure \ref{fig:temp}c:
The $T^{\rm{N{_2}H^{+}}}_{\rm ex}$ is $>5$ K, which increases to $\sim11$ K in the dense regions where the young cluster is located. 
The temperature around the young cluster is also high in the $T_{\rm dust}$ map.

The temperature distributions of $T^{\rm{co}}_{\rm ex}$, $T_{\rm dust}$, and $T^{\rm{N{_2}H^{+}}}_{\rm ex}$ roughly show a similar tendency: All of the maps show a trend of decreasing temperature with increasing distance from the \Hii region, 
suggesting that the molecular gas and dust in W40 are warmed by the radiation from the OB cluster.


\subsection{The velocity structure} 

To investigate the global velocity structures, we made mean velocity $V_{0}$ map of the \eco, CCS, and \nnh emission lines.
For \eco and CCS, $T_{\rm mb}$ was measured through fitting the lines with a single Gaussian at each observed position.
${V_0}$ was then measured at each position as
\begin{equation}
\label{eq:tex}
{V_0}=\frac{{\int T_{\rm mb}v\,dv}}{{\int T_{\rm mb}\,dv}}\,.
\end{equation}
We show the results in figures \ref{fig:V0map}a and \ref{fig:V0map}b.
For the N$_2$H$^{+}$ ($J=1-0$) line in figure \ref{fig:V0map}c, we show the values of $V_{0}$
fitted with equations (\ref{eq:radiative}) and (\ref{eq:nnh})
as described in the previous subsection.

The global spatial distributions of ${V_0}$ of the three emission lines show a similar
velocity gradient from inner regions of W40 to the Serpens South region.
We found that Serpens South is accompanied by molecular gas with multiple velocity components, 
as reported in the previous studies for limited regions \cite[e.g.,][]{Kirk2013, Fernandez2014,Nakamura2014}:
In the southern part of Serpens South, the three emission lines are detected at $\sim6-7$ km s$^{-1}$, while
these are detected at higher velocities in the northern part.

In figure \ref{fig:dV}, we also show the velocity dispersion ${\Delta V}$ maps for the three emission lines.
The ${\Delta V}$ maps for the \eco and CCS emission lines are derived from a single Gaussian fitting for each line profile.
In the case of \eco, ${\Delta V}$ is especially larger around the \Hii region. 
The reason for the larger ${\Delta V}$ is that there are multiple velocity components as can be seen in the spectra in the figure.
We also found that ${\Delta V}$ of CCS and \nnh tends to be large around the young cluster of Serpens South, which is also seen in the spectra shown in figure \ref{fig:spectra}c.
This is also thought to be due to the existence of multiple velocity components, as we will suggest in subsection \ref{sec:four velocity}.


\subsection{The distribution of the fractional abundance}

To understand chemical characteristics of the observed region,
we estimated fractional abundances of \eco, CCS, and \nnh.
On the assumption of the LTE condition, 
the column density of \eco and CCS, $N$(\eco) and $N$(CCS), respectively, can be approximated
using the integrated intensity of the lines, $\int T{\rm_{mb}}dv$, with the following formula \cite[e.g.,][]{Hirahara,Mangum2015},

\begin{equation}
\label{eq:column1}
N=\frac{3h}{8\pi^{3}}\frac{Q}{\mu^{2}S_{ij}}\frac{e^{Eu/k{T}_{\mathrm{ex}}}}{e^{T_{0}/T_{\mathrm{ex}}}-1}\frac{\int \beta^{-1}T{\rm_{mb}}dv}{J(T_{\mathrm{ex}})-J(T\mathrm{_{bg}})}
\end{equation}
where 
$Q$ is the partition function approximated as $Q=k\,T_{\rm ex}/h\,B_{0}+1/3$ \citep[e.g.,][]{Mangum2015}. 
$B_{0}$ is the rotational constant of the molecule. 
$\mu$ is the dipole moment, $E_{u}$ is the energy of the upper level, and $S_{ij}$ is the intrinsic line strength of the transition for $i$ to $j$ state. 
$\beta$ is the escape probability related with the optical depth $\tau$ as $\beta =(1 - {e^{ - \tau}})/\tau$,
and $\beta=1$ for $\tau\ll 1$. 
For $T_{\rm ex}$, we adopted $T^{\rm{co}}_{\rm ex}$ for \eco, and a fixed value of 5 K for CCS measured in dark clouds \citep[e.g.,][]{Hirota2009}.
The spectral line parameters are taken from Splatalogue.  

For the estimation of the column density of \nnh, $N$(\nnh), we used the following formula \cite[e.g.,][]{Caselli2002,Mangum2015},
\begin{equation}
\label{eq:column2}
N({\rm N}_{2}{\rm H}^{+})=\frac{8\pi^{3/2}}{2\sqrt{\rm{ln}2}}\frac{Q {\nu^{3}}}{c^{3}A g_{u}}\frac{\tau_{\rm tot}\Delta V}{1-e^{-{T_{0}/T_{\mathrm{ex}}}}} \, 
\end{equation}
where $g_{u}$ is the degeneracy of the upper level of a transition, 
$A$ is the Einstein coefficient for spontaneous emission.
We calculated $Q$ using the excitation temperature $T_{\rm ex}^{\rm N_2H^+}$
derived in Section \ref{sec:temperature}.
We summarize the line parameters used in this study in table \ref{tab:parameter}.

We then estimated the fractional abundances of the three molecules, $f$(\eco), $f$(\nnh), and $f$(CCS), by dividing their derived column densities by the H$_2$ column density derived from the dust data by $\it{Herschel}$.
For this, we re-gridded and smoothed the H$_2$ column density data to the same angular resolution of the 45 m data ($\sim22\arcsec$).
Figure \ref{fig:fmap} shows the resultant fractional abundance maps.

The value of $f$(\eco) reaches as high as $\sim10^{-6}$ around the center of W40,
but it is apparently much lower 
in Serpens South, which may be due to the depletion of \eco onto dust in the cold and dense interior of the Serpens South filament.
On the other hand, $f$(\nnh) is high in Serpens South.
$f$(CCS) is also high in Serpens South, which means that the Serpens South region is in an earlier evolutionary stage than the W40 region \citep[e.g.,][]{Shimoikura2012,Shimoikura2018}.
In the Serpens South region, the low values of $f$(\eco), $f$(CCS), and $f$(\nnh) are seen at the position of the young cluster.
Because the difference of the fractional abundances infer the difference of chemical reaction time
\citep[e.g.,][]{Suzuki1992}, the northern part of the Serpens South filament is likely to be
younger than the southern part. The next cluster formation may occur in the northern part 
of the filament, as suggested by \cite{Nakamura2014}.


\subsection{The four velocity components}\label{sec:four velocity}

To investigate the velocity distribution of the C$^{18}$O emission in more detail
and to investigate the relation between the ionization front and the molecular gas,
we made channel maps of the C$^{18}$O emission with a step of 0.3 km s$^{-1}$, and show them in figure \ref{fig:channel1}
together with the $Spitzer$ composite image same as figure \ref{fig:spitzer}.

In the figure, the C$^{18}$O distribution shows a complex velocity structure and varies from panel to panel.
Most of the C$^{18}$O emission concentrates at $\sim$ 7 km s$^{-1}$, and the emission associated
with Serpens South is also found around this velocity.
At low velocities in panels a $-$ e, the C$^{18}$O emission is found only around the OB cluster.
At higher velocities in the other panels, the emission tends to be located in the periphery of the \Hii region seen in the
$\it{Spitzer}$ image, for which
we suggest that the C$^{18}$O emission should trace the gas swept up by the expansion of the \Hii region.

As shown in the panels $1-3$ of figure \ref{fig:dV}, there found a number of velocity components detected in C$^{18}$O in the observed region.
Based on the inspection of the channel maps, we decided to categorize the emission into
four groups in terms of the apparent spatial distributions around the \Hii region.
We shall call them  5, 6, 7, and 8 {km s$^{-1}$ components according to their
typical LSR velocities. 
These components show different features and are helpful to understand the global velocity
structure of the observed region.
Figure \ref{fig:4comp1} shows distributions of the four components
superposed on the $\it{Spitzer}$ image (figure \ref{fig:spitzer}).
In the following, we summarize the characteristics of the components seen in figures \ref{fig:V0map} $-$ \ref{fig:4comp1}.

{\bf 5 km s$^{-1}$ component}:
As seen in figure \ref{fig:4comp1}a, The molecular gas with LSR velocities 
in the range $4.3 \le V_{\rm LSR} \le 5.9 $ km s$^{-1}$
is mainly located just around the \Hii region. 
The excitation temperature of this component is very high in the \Hii region, 
suggesting that the component is located near the exciting sources (the OB cluster) and is warmed by the radiation from them.

{\bf 6 km s$^{-1}$ component}:
As seen in figure \ref{fig:4comp1}b, the molecular gas with LSR velocities 
$5.9 \le V_{\rm LSR} \le 6.8 $ km s$^{-1}$  are found around the \Hii region
and at the southern part of the main body of Serpens South.
The component located around the \Hii region shows a ring-like structure
surrounding the 5 km s$^{-1}$ component.
In addition, there is a filament at $\sim6$ km s$^{-1}$ with an elongation of $\sim3$ pc
at the most western side in the observed area (as labelled ``filament A").
This filament shows an anti-correlation with the 8 km s$^{-1}$ component shown in figure \ref{fig:4comp1}d.

{\bf 7 km s$^{-1}$ component}:
As seen in figure \ref{fig:4comp1}c, the molecular gas with LSR
velocities in the range $6.8 \lesssim V_{\rm LSR} \lesssim 7.5 $ km s$^{-1}$ 
are found mainly around the Serpens South region as well as in the periphery of the \Hii region.
The diffuse and extended emission is also detected in the \Hii region, being coincident with the dark lane.
The component associated with the dark lane is extending from the Serpens South region to the W40 region,
as traced by the broken line.

{\bf 8 km s$^{-1}$ component}:
As seen in figure \ref{fig:4comp1}d, the molecular gas with LSR velocities in the range $7.5 \lesssim V_{\rm LSR} \lesssim 8.8 $ km s$^{-1}$  
appears to lie at the boundary of the \Hii region.
The component is seen mostly in the north-western side of the \Hii region. \\

In our previous study using the $^{12}$CO ($J=3-2$) and HCO$^{+}$ ($J=4-3$) emission lines performed only in the limited region around the W40 \Hii region, we reported that there are velocity components at $V_{\rm LSR} \simeq 3$, 5, 7, and 10 km s$^{-1}$ \citep{Shimoikura2015}. 
These velocity components are mostly single peaked components with well-defined
radial velocities and are not categorized in the same way as the four components in the above,
but the 5 and 7 km s$^{-1}$ components found in the previous study can be merged into those categorized in this work.
The $\sim3$ and $\sim10$ km s$^{-1}$ components found in the previous work are not detected in \eco in this study with a good signal-to-noise ratio. However, they are clearly detected in the $^{12}$CO ($J=1-0$) and
$^{13}$CO ($J=1-0$) lines.
In figure \ref{fig:2comp}, we show the distributions of the $\sim3$ and $\sim10$ km s$^{-1}$
components detected in the $^{13}$CO ($J=1-0$) data, and we shall call them the 3 and 10 km s$^{-1}$ component, respectively.
These components are seen around the OB cluster as we already found in the earlier study.
In this study, we further found that the 10 km s$^{-1}$ component is widely distributed over the observed area, especially
in the periphery of the \Hii region like the 8 km s$^{-1}$ component in figure \ref{fig:4comp1}d.

In summary, the distributions of the identified components indicate an interaction between the \Hii region and the surrounding
molecular gas. The components at higher velocities (i.e., 7, 8, and 10 km s$^{-1}$)
tend to exhibit an arc-shaped structure
and are located at the boundary of the \Hii region, suggesting that they may be tracing expanding shell
of the \Hii region.


\subsection{ The two expanding shells }

To better understand the spatial and velocity distributions of molecular gas and to investigate
the effect of the expanding \Hii region on the surroundings, we made position-velocity (PV) diagrams
across the observed region using the \eco data.
Figure \ref{fig:PV1} displays a series of the PV diagrams measured along the cuts
centered at the position of IRS1A South toward various directions.
In each PV diagram, the position of IRS1A South as well as the positions
separated by $\pm 1000\arcsec$ away from the source roughly corresponding to the boundary
of the \Hii region are indicated by the white broken lines.

We found elliptical structures in the diagrams as
we indicate two of them by ellipses in panels i and p where the structures are evident.
The elliptical structures should be due to expanding shell(s) of the \Hii region centered at IRS1A South or the OB cluster.
We suggest that there are two shell-like structures: One is the small inner shell just around IRS1A South
shown by the ellipse in panel p, and the other is the large outer shell corresponding to the boundary of the \Hii region
delineated by the ellipse in panel i.
The ellipses in the panels indicate that the inner and outer shells have a radius of $\sim0.5$ pc and $\sim2.5$ pc,
and an expanding velocity of $\sim3$ km s$^{-1}$.
The expansion time scale of the two shells can be estimated by dividing the radius by the velocity
to be $1.6\times10^{5}$ yr and $8.1\times10^{5}$ yr for the inner and outer shells, respectively,
which is consistent with the dynamical age of the \Hii region ($1.9\times10^{5}-7.8\times10^{5}$ yr)
estimated by \cite{Mallick}
based on the radio continuum observations.
We also estimated the mass within the outer shell (the northwest side of the W40 region) to be $\sim1 \times 10^{3} M_{\odot}$\footnote{The mass is derived from the $\it{Herschel}$-based $N$(H$_2$) data as $M=\alpha m_{\rm H} \int_S N({\rm H_2})dS$ 
where $S$ is the area of the \Hii region determined within the radius of the outer shell ($R$ = 2 pc),
$\alpha$ is the mean molecular weight taken to be 2.8, and $m_{\rm H}$ is the mass of a hydrogen atom.}.

We suggest a possibility that the two shell-like structures were created
due to the inhomogeneous density distribution of the dense gas around the \Hii region.
As seen in figure \ref{fig:4comp1}, there are some patchy dense clumps (e.g., the $5 - 7$ km s$^{-1}$ components)
in the vicinity of IRS 1 A South and the OB cluster. Expansion of the \Hii region should be
slowed or blocked by such clumps,
and some parts of the expanding shell facing to the clumps should appear as the inner shell,
while the rest of the shell should appear as the outer shell, as illustrated in figure \ref{fig:hourglass}.
We further suggest that this may be the very mechanism to
create the hourglass-shaped structure of the \Hii region.


\subsection{ The cluster formation in Serpens South }

We investigated whether the expansion of the \Hii region influenced the cluster formation in Serpens South.
Figure \ref{fig:PV2}a (same as figure \ref{fig:PV1}b) shows the PV diagram measured along the cut b'-b
crossing the position of the young cluster of Serpens South.
The outer shell delineated by the ellipse in figure \ref{fig:PV2}a is not seen very well in
the \eco emission, so we show in figures \ref{fig:PV2}b and c the PV diagrams of the
$^{13}$CO and $^{12}$CO  emission taken along the same cut, respectively, where
the outer shell can be better recognized.
As seen in the PV diagrams, the dense gas associated with the young cluster (evident in figure \ref{fig:PV2}a) is very likely
to be located at the surface of the outer shell, indicating that the natal clump of the cluster can be affected by the shell.
Actually, as seen in figure \ref{fig:PV2}a, the velocity dispersion is enhanced abruptly at the position of the cluster (up to $\sim 2$ km s$^{-1}$)
compared to the adjacent positions, which can be recognized also in the ${\Delta V}$ maps presented in figures \ref{fig:dV}b and c.
Though the observed enhancement of the velocity dispersion may be partially due to the feedback of star formation (e.g., outflows),
this may support the idea that the natal clump of the young cluster is interacting with the outer shell
being compressed by the expansion of the shell, which may have induced the formation of the young cluster.

To further investigate the influence of the outer shell, 
we created a PV diagram for the C$^{18}$O and CCS emission
along the major axis of the Serpens South filament, which is shown in figure \ref{fig:PV2}d.
In the figure, there can be found two major velocity components as labeled ``X" and ``Y"
which correspond to subfilaments extending to the southeast and northwest, respectively.
The subfilament X has a velocity ($\sim 6$ km s$^{-1}$) slightly lower than the
systemic velocity ($\sim 7$ km s$^{-1}$), and the subfilament Y has a velocity same as the systemic velocity.
The young cluster is apparently located at the intersection of the subfilaments X and Y,
and both of them are located around the surface of the outer shell.
This positional coincidence infers a scenario that the subfilament X has been accelerated by the
expansion of the outer shell and recently collided against the subfilament Y to induce cluster
formation at the intersection.

The previous studies \cite[e.g.,][]{Kirk2013,Nakamura2014} also reported that Serpens South consists of
molecular filaments with different velocities.
\cite{Nakamura2014} suggested that the collisions of the filaments with the different radial velocities may
have triggered the cluster formation like what we discuss in the present study.
The PV diagram of figure \ref{fig:PV2}d shows a velocity gradient from southeast
to northwest of $\sim0.7$ km s$^{-1}$ pc$^{-1}$, which is consistent with
that found by the previous studies \citep{Kirk2013,Fernandez2014}.
\cite{Kirk2013} interpreted that the velocity gradient is an evidence of material flowing
inside the filaments, leading to infall toward the intersection of the filaments where the cluster formation takes place. 
However, \cite{Fernandez2014} suggest that the velocity gradients can also occur in scenarios where filaments are
created by large-scale turbulence.

We basically agree with the scenario suggested by \citet{Nakamura2014} that the cluster formation
was induced by collisions of filaments. Our subfilaments X and Y correspond to their filaments
F8 and F1-3, respectively (see their figure 3).
We further suggest that the outer shell of the W40 \Hii region have played a crucial role
to accelerate the southeastern part of the filament (i.e., the subfilament X)
leading to the collisions of the filaments. Also, the outer shell may have influenced directly
on the natal clump of the cluster by compressing it to trigger the cluster formation.
 


\subsection{ The global 3D structure }
In our previous work \citep{Shimoikura2015}, we investigated the three-dimensional structure of W40 just around the \Hii region. 
Because we are now almost confident that W40 and Serpens South belong to the same system, we will try to investigate the three-dimensional structure of the entire system based on the data obtained in this study.

In figure \ref{fig:4comp2}, we present the spatial distributions of each component together with the IR images.
Considering the velocity difference of each component with respect to the systemic velocity at 7 km s$^{-1}$
as well as the expanding motion of the \Hii region,
the W40 + Serpens South system should be structured basically in such a way that
the 5 km s$^{-1}$ and  6 km s$^{-1}$ components are located on the near side of the \Hii region
facing toward us, and the 8 km s$^{-1}$ component is located on the far side.

Based on the above picture, we attempt to reveal the locations of the individual components in the system.
The 5 km s$^{-1}$ component is distributed only around the center of the \Hii region.
It must be the dense gas close to and being blown away by the \Hii region, and moving toward us.
This component is probably located at the inner shell.
The 6 km s$^{-1}$ component and a part of the 7 km s$^{-1}$ component seen around the main body of Serpens South
appear as the absorption (i.e., dark) in the $\it{Spitzer}$ image, delineating the shape of the Serpens South filaments.
The absorption indicates that they must be the dense gas located in the foreground of the \Hii region.
The dark lane located in the waist of the \Hii region at 7 km s$^{-1}$ 
also obscures the central part of the \Hii region in the $\it{Spitzer}$ image,
and therefore it must be located in the foreground.
The 8 km s$^{-1}$ component is found mainly in the northern part of the observed region,
and it does not appear as the absorption in the $\it{Spitzer}$ image, suggesting
that it is located rather in the back of the \Hii region. 

We further attempt to model the locations of the 3 and 10 km s$^{-1}$ components
identified with the $^{13}$CO data in the same way.
We found a high temperature of $\sim30$ K for the components as seen in figure \ref{fig:temp}a, 
suggesting that they are located close to the central OB stars.  
As seen in the PV diagram of figure \ref{fig:PV2}c,  
the 10 km s$^{-1}$ component is apparently located at the inner shell of the \Hii region
on the far side to the observer,
which is consistent with the suggestion in our previous study \citep{Shimoikura2015}.
The 3 km s$^{-1}$ component is also likely to be located at the inner shell but on the near side to the observer,
like in the case of the 5 km s$^{-1}$ component.

In figure \ref{fig:model}, we summarize the geometry of the entire W40 and Serpens South system
inferred from the above discussion.
The morphology and kinematics around the W40 \Hii region are consistent with what we proposed before \citep{Shimoikura2015}.
Finally, using the $\it{Herschel}$-based $N$(H$_2$) data, we estimated the total mass of the four components to be
$\sim200$, $\sim1200$, $\sim5000$, and $\sim2000 M_{\odot}$ for the 5, 6, 7, and 8 km s$^{-1}$ component, respectively\footnote{
We defined the extents of the four components by the 1.0 K km s$^{-1}$ contours of the \eco intensity (e.g., figure\ref{fig:4comp1}),
and derived the masses of the components by integrating $N$(H$_2$) within the extents. For a pixel belonging to
two or more velocity components, the mass in the pixel is shared by the components proportionally to the \eco
intensities of the components.}.

As seen in the PV diagrams (figure \ref{fig:PV1}), 
the velocity distribution of each component is narrow with a line width of $\sim1$ km s$^{-1}$, except for the 5 km s$^{-1}$ component. 
According to \cite{Beaumont} who observed 43 bubbles associated with \Hii regions, there is often a ring of cold gas
driven by massive stars, and \Hii regions form in flat molecular clouds with a thickness of a few parsecs.
We point out a possibility that some
filamentary structures seen around the equator of the hourglass-shaped W40 \Hii region can be
an edge-on view of a sheet-like or ring-like cloud, like what was suggested by \cite{Beaumont}
for other \Hii regions.

In general, as an advancing ionization front wraps pre-existing clumps, its pressure triggers a gravitational collapse and star formation \cite[e.g.,][]{Elmegreen}.
The two shells we found in this study have apparently been created by the \Hii region, and they
may be such a place of sequential star formation, as it is actually evident in figure \ref{fig:YSO} showing
the distributions of YSOs \citep{Andre2010, Konyves2015}.
In the W40 and Serpens South system, star formation probably occurred in W40 first, and then the next-generation stars
are forming in clouds along the expanding shells.
We suggest that the young cluster in Serpens South was also induced by the expansion of the \Hii region.
This idea is consistent with the fact that the CCS emission is strongly detected in the Serpens South region while it is not detected in W40, indicating that the Serpens South region is much younger than the W40 region.


\section{ CONCLUSIONS }\label{sec:conclusions}

Large-scale mapping observations of the star-forming regions of W40 and Serpens South were performed in $^{12}$CO, $^{13}$CO, C$^{18}$O, CCS, and \nnh using the NRO 45 m telescope. 
We revealed the complex distributions of the molecular emission lines and their radial velocities in the regions.
We found that the C$^{18}$O emission is smoothly distributed over the both W40 and Serpens South regions,
suggesting that the two regions belong to the same system.
The CCS and \nnh emission lines are enhanced in Serpens South.
Serpens South is characterized by its filamentary structure in the infrared images such as by $Spitzer$ and $Hershell$,
and it is accompanied by a young cluster near the center of the structure. 
On the other hand, in W40, the CCS emission is not detected significantly, and the \nnh emission
is detected almost only around an OB cluster at the center.

Based on the C$^{18}$O observations, we divided the molecular gas into four velocity components
which we call 5, 6, 7, and 8 km s$^{-1}$ components according to their
typical LSR velocities.
We found that two elliptical structures are seen in the position-velocity diagrams,
which can be explained as a part of two expanding shells.
One of these shells is the small inner shell found in the vicinity of the \Hii region,
and the other is the larger outer shell corresponding to the boundary of the \Hii region. 
The dense gas associated with the young cluster in Serpens South is likely to be interacting with
the outer shell, for which we suggest that the cluster was induced to form by the interaction with the expanding
outer shell.

Strong CCS emission is detected in Serpens South while it is not detected in W40,
which means that the Serpens South region is
younger than the W40 region in terms of chemical reaction. 
This result is consistent with the scenario that the star formation initially occurred in W40
and then the young cluster in Serpens South was induced to form by the interaction with the
expanding shell of the \Hii region.

Based on the morphologies and velocity distributions of the four velocity components
as well as their appearance in the infrared images, we made a three-dimensional geometrical
model of the W40 and Serpens South regions.


\begin{ack}
We are grateful to the referee for providing useful comments to improve this paper.
We thank Yoshiko Hatano, Akifumi Yamabi, Sho Katakura, and Asha Hirose for their support of the observations.
We are also grateful to the other members of Star Formation Legacy project.
This work was supported by JSPS KAKENHI Grant Numbers JP17K00963, JP17H02863, JP17H01118, JP16K12749. 
YS received support from the ANR (project NIKA2SKY, grant agreement ANR-15-CE31-0017).
The 45 m radio telescope is operated by NRO, a branch of NAOJ. 
This research has made use of data from the Herschel Gould Belt survey (HGBS) project (http://gouldbelt-herschel.cea.fr). 
The HGBS is a Herschel Key Programme jointly carried out by SPIRE Specialist Astronomy Group 3 (SAG 3), scientists of several institutes in the PACS Consortium (CEA Saclay, INAF-IFSI Rome and INAF-Arcetri, KU Leuven, MPIA Heidelberg), and scientists of the Herschel Science Center (HSC).
\end{ack}




\clearpage

\begin{table}
\caption{The observed molecular lines} 
\begin{tabular}{lcrcc}  
\hline\noalign{\vskip3pt} 
	\multicolumn{1}{c}{Molecule} & \multicolumn{1}{c}{Transition} & \multicolumn{1}{c}{Frequency{\footnotemark[$1$]}} & \multicolumn{1}{c}{Effective beam size} & \multicolumn{1}{c}{$\Delta T_{\rm mb}${\footnotemark[$2$]} }\\
	\multicolumn{1}{c}{} & \multicolumn{1}{c}{} & \multicolumn{1}{c}{(GHz)} & \multicolumn{1}{c}{($\arcsec$)} & \multicolumn{1}{c}{(K)} \\  [2pt] 
\hline\noalign{\vskip3pt} 
N$_2$H$^{+}$	&	$J=1-0$				&	  93.1737637 	&	$24$	&	0.27	\\
CCS			&	$J_{N}$=8$_{7}-7_{6}$	&	  93.8700980 	&	$24$	&	0.26	\\
C$^{18}$O	&	$J=1-0$				&	109.7821760 	&	$22$	&	0.33	\\
$^{13}$CO	&	$J=1-0$				&	110.2013540 	&	$22$	&	0.36	\\
$^{12}$CO	&	$J=1-0$				&	115.2712040 	&	$22$	&	0.87	\\
\hline
\end{tabular} \label{tab:line}
   \begin{tabnote}
{\footnotemark[$1$]} The rest frequency for the N$_2$H$^{+}$ line is taken from \cite{Pagani2009}, and those of the other lines are taken from the website of \cite{Lovas}.\\
{\footnotemark[$2$]} The one sigma noise level at a velocity resolution of 0.1 km s$^{-1}$.
   \end{tabnote}
\end{table} 


\begin{table}
\caption{Constants of the observed molecular lines} 
\begin{tabular}{lrrrrrcc} 
\hline\noalign{\vskip3pt} 
\multicolumn{1}{c}{Molecule} & \multicolumn{1}{c}{Transition} & \multicolumn{1}{c}{$S_{ij}$} & \multicolumn{1}{c}{$B_{0}$}  & \multicolumn{1}{c}{$\mu$} & \multicolumn{1}{c}{$E_u$} & \multicolumn{1}{c}{$g_u$} & \multicolumn{1}{c}{$A$}\\
\multicolumn{1}{c}{} & \multicolumn{1}{c}{} & \multicolumn{1}{c}{} & \multicolumn{1}{c}{(GHz)}  & \multicolumn{1}{c}{(Debye)} & \multicolumn{1}{c}{(cm$^{-1}$)} & \multicolumn{1}{c}{} & \multicolumn{1}{c}{($s^{-1}$)}\\[2pt] 
\hline\noalign{\vskip3pt} 
 N$_2$H$^{+}$	&	$J=1-0$				&	-	&	46.586867	&	-	&	-	&	3 		&	3.628$\times$10$^{-5}$ \footnotemark[$*$]	\\
CCS			&	$J_{N}$=8$_{7}-7_{6}$	&	7.97		&	6.47775036	&	2.88		&	13.82557	&	-	&	-	\\
C$^{18}$O	&	$J=1-0$				&	1.00 		&	54.8914 		&	0.110 	&	3.66194 	&	-	&	-	\\
\hline
\end{tabular} \label{tab:parameter}
   \begin{tabnote}
{\footnotemark[$*$]}\cite{Pagani2009}
\end{tabnote}
\end{table} 



\begin{figure*}
\begin{center}
\includegraphics[scale=.4]{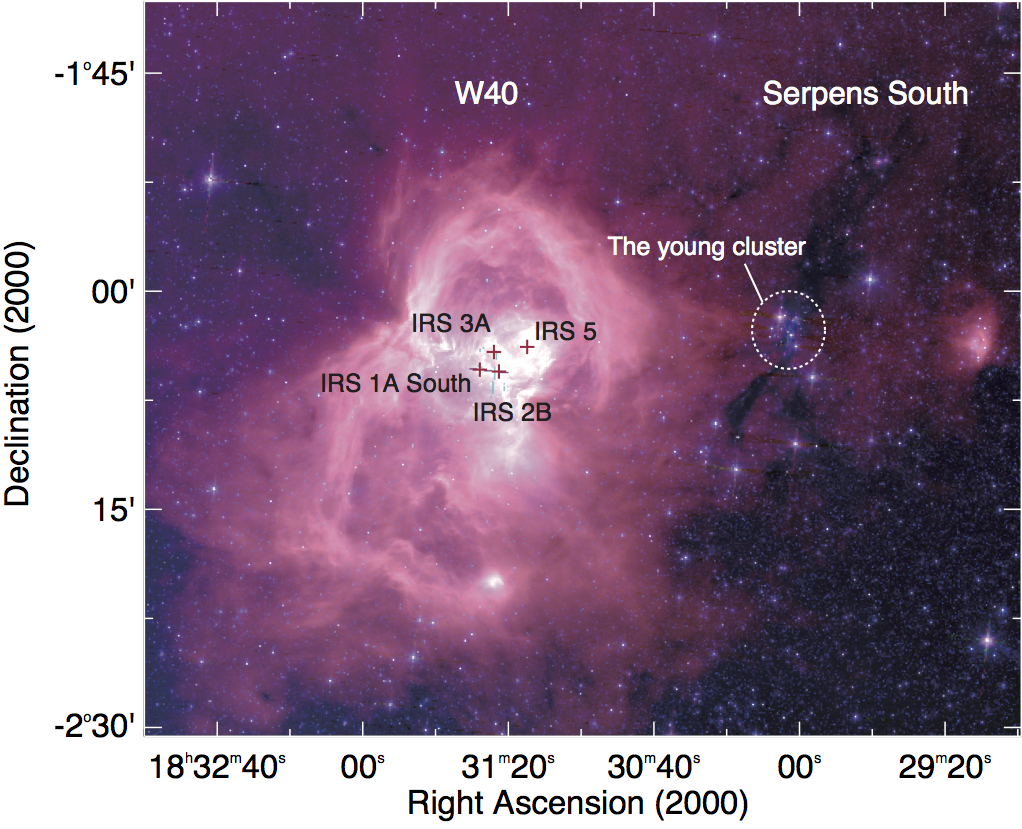}
\end{center}
\caption{
The W40 and Serpens South regions shown in a three-color composite image made with the 
$8.0 \micron$ (red), $5.8 \micron$ (green) and $4.5 \micron$ (blue) data by $\it{Spitzer}$.
The plus signs indicate the positions of the high mass infrared sources IRS 1A South, IRS 2B, IRS 3A, and IRS 5
\citep{Smith, Shuping2012}. 
\label{fig:spitzer}}
\end{figure*}


\begin{figure*}
\begin{center}
\end{center}
\includegraphics[scale=.5]{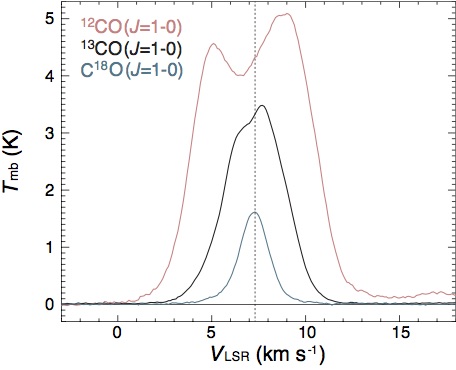}
\caption{
The $^{12}$CO, $^{13}$CO, and \eco  spectra averaged over the pixels detected at the $>5 \sigma$ noise level.
The vertical dashed line indicates the peak radial velocity of the averaged \eco spectrum ($V_{\rm LSR}=7.3$ km s$^{-1}$).
\label{fig:spe}}
\end{figure*}


\begin{figure*}
\begin{center}
\includegraphics[scale=.4]{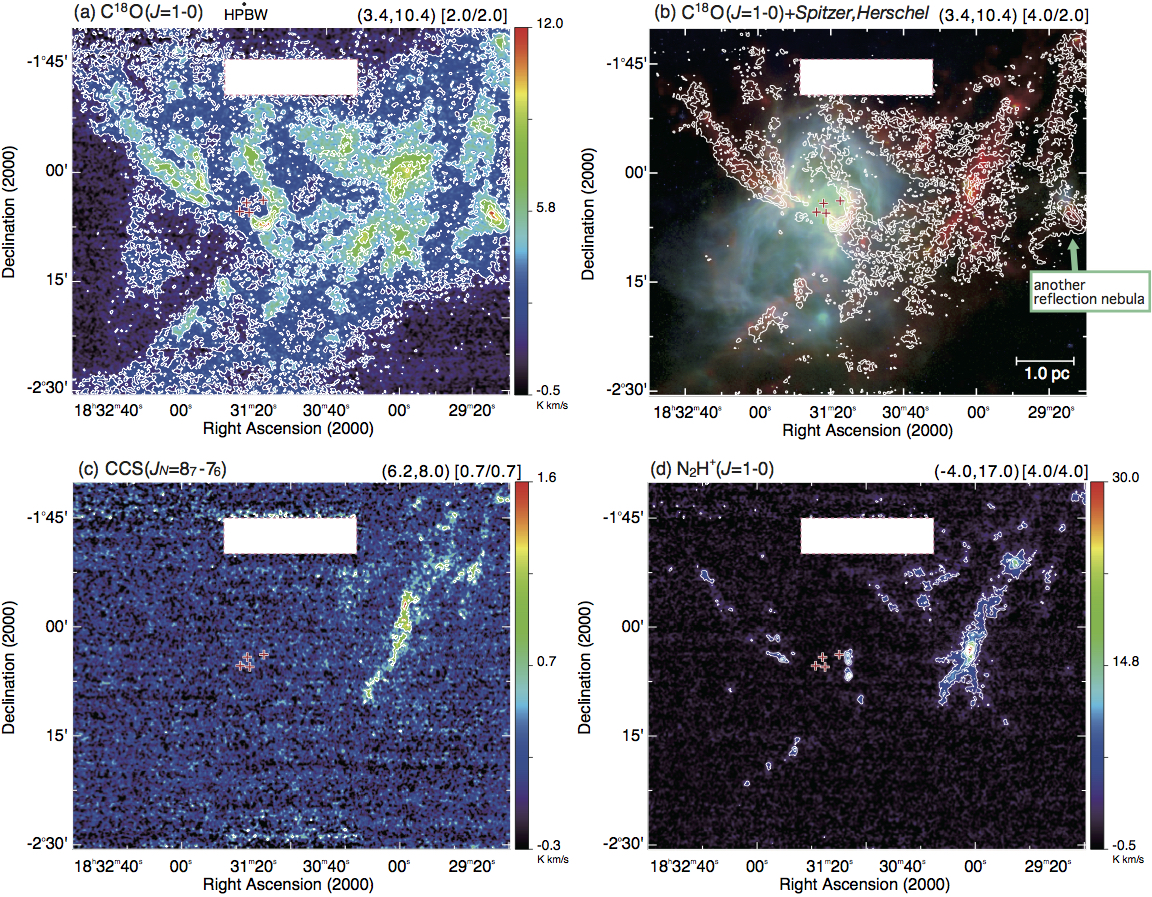}
\end{center}
\caption{
Velocity-integrated intensity maps of the (a) C$^{18}$O, (c) CCS, and (d) \nnh emission lines, respectively. 
Velocity ranges in units of km s$^{-1}$ used for the integration
are shown in the parentheses, and the contour levels of
the lowest / intervals
in units of K km s$^{-1}$ are shown in the brackets
above each panel.
In panel (b), the \eco map (white contours) is overlaid on a composite color image of
$Herschel$ 500 \micron \,(red), 70 \micron \,(green), 
and $Spitzer$ 8.0 \micron \,(blue). 
\label{fig:iimap}}
\end{figure*}

\begin{figure*}
\begin{center}
\includegraphics[scale=0.5]{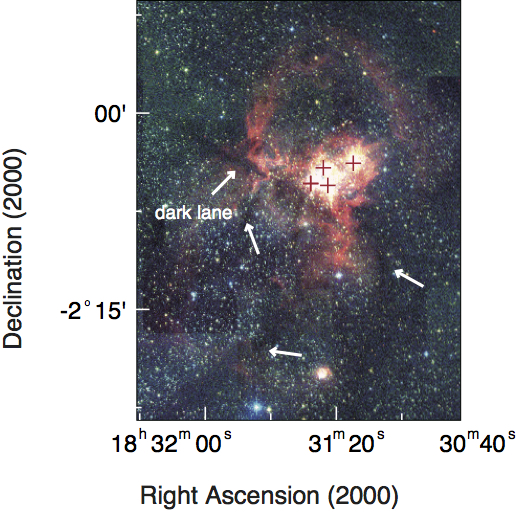}
\end{center}
\caption{
Composite image of W40 made with $\it{Spitzer}$ $8\micron$ (red), 2MASS $K$s (green), and $H$ (blue).
Some dark lanes are shown with a white arrow.
\label{fig:dark}}
\end{figure*}

\begin{figure*}
\begin{center}
\includegraphics[scale=0.4]{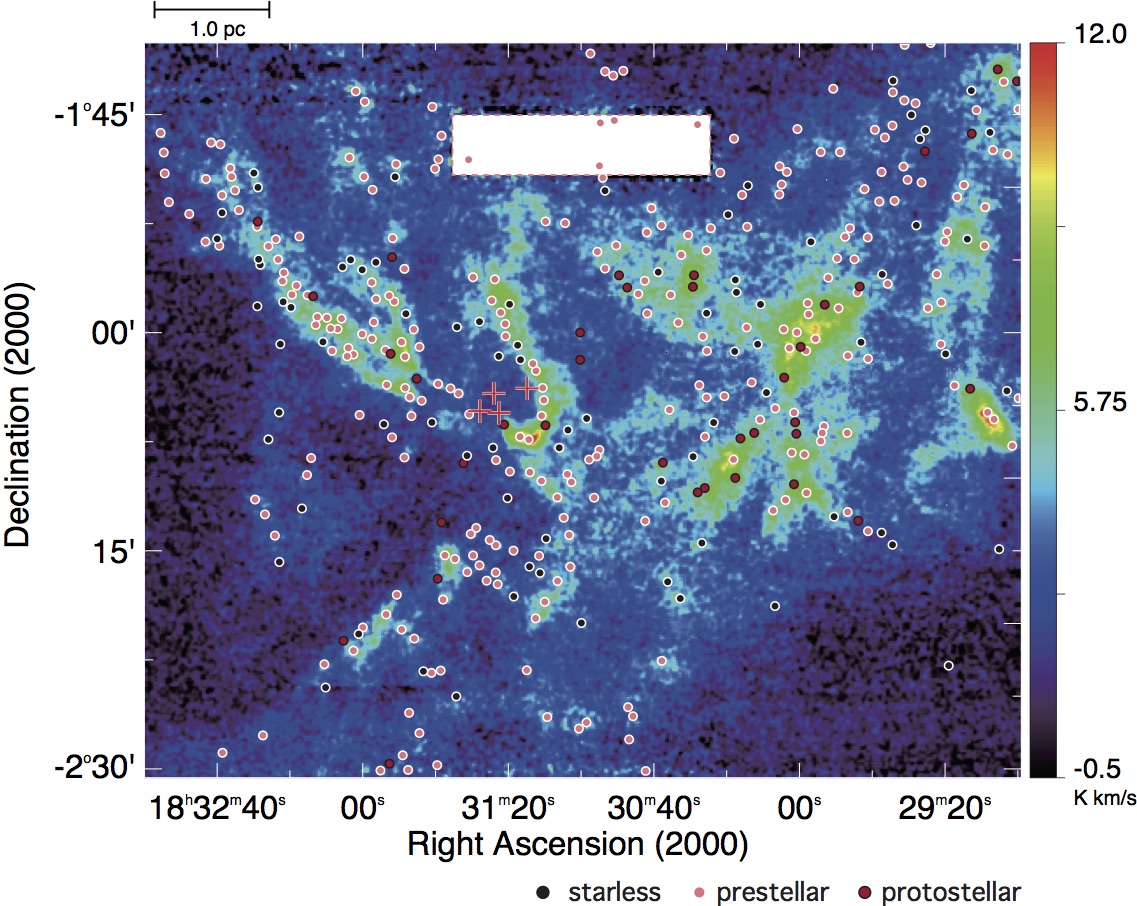}
\end{center}
\caption{
Distributions of cores reported by \citet{Andre2010} and \citet{Konyves2015} shown on the \eco map.
Starless, prestellar, and protostellar cores are indicated by black, pink, and red circles, respectively.
The starless cores are the cores without YSO candidates and the prestellar cores are the cores with YSO candidates \cite[see,][]{Konyves2015}.
\label{fig:YSO}}
\end{figure*}

\begin{figure*}
\begin{center}
\includegraphics[scale=.4]{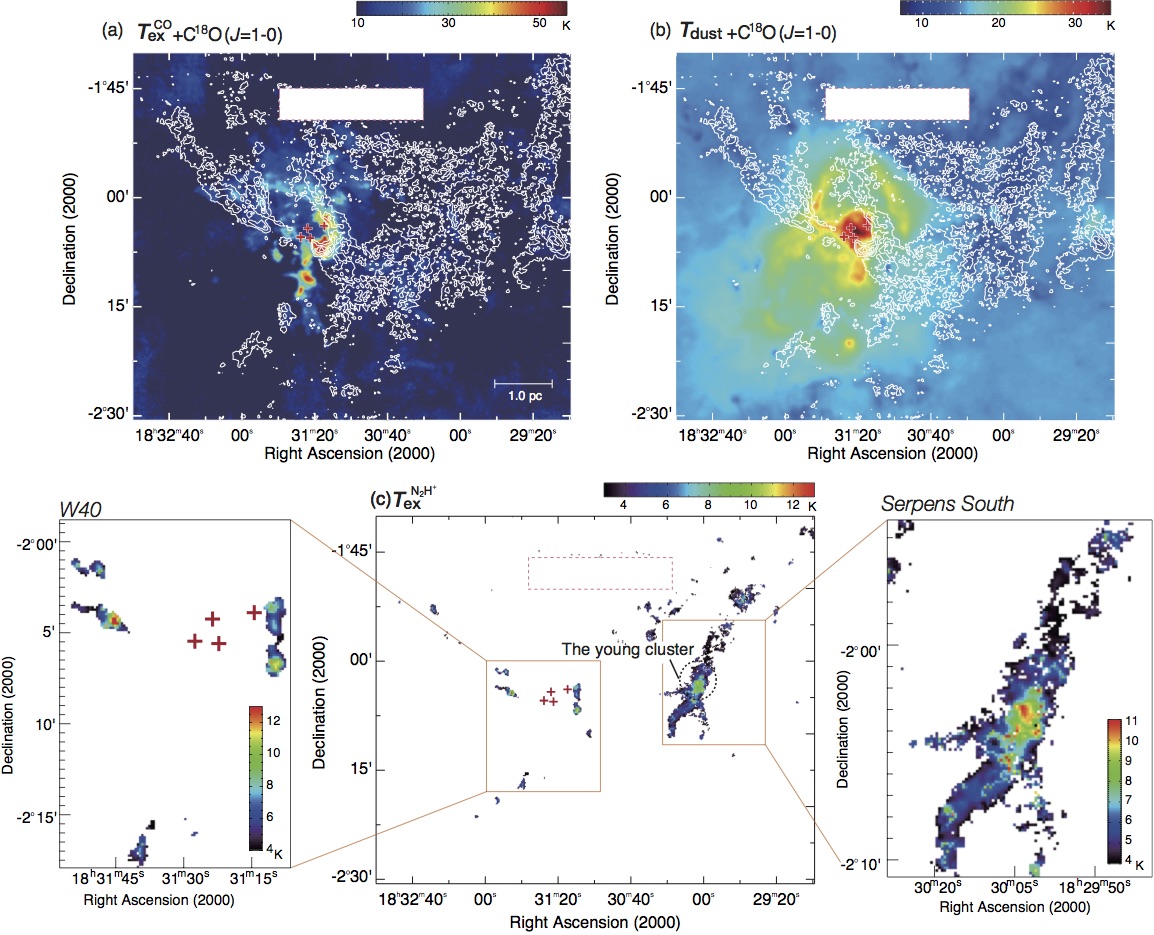}
\end{center}
\caption{
(a) Excitation temperature map derived from the $^{12}$CO ($J=1-0$) data.
(b) Dust temperature map \citep{Andre2010, Konyves2015}. 
(c) Excitation temperature map of N$_2$H$^{+}$ ($J=1-0$) (middle),
and the close-up view of the W40 region (left) and the Serpens South region (right).
The \eco intensity map in figure \ref{fig:iimap}a is overlaid in panels (a) and (b)
by the white contours. The lowest contour and the contour interval are 4 K km s$^{-1}$ and 2 K km s$^{-1}$.
\label{fig:temp}}
\end{figure*}

\begin{figure*}
\begin{center}
\includegraphics[scale=.4]{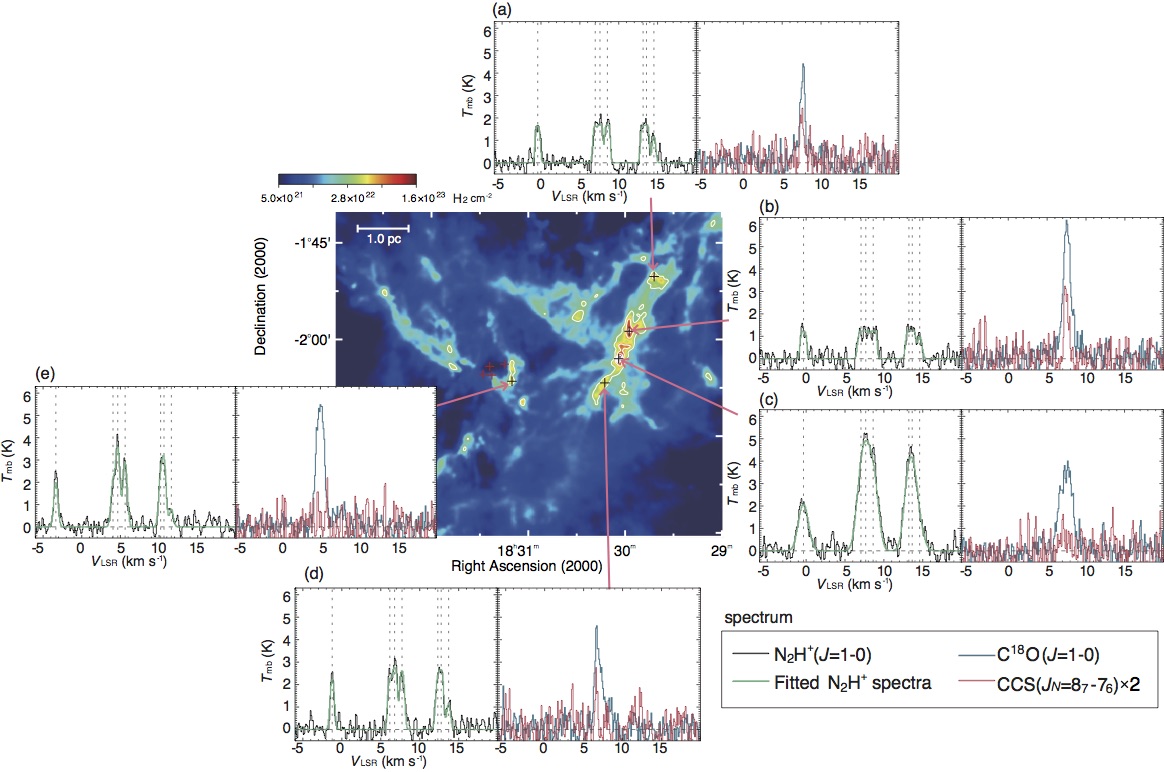}
\end{center}
\caption{
\eco, \nnh, and CCS spectra observed at some positions indicated by the black plus signs. 
Results of the hyperfine fit of the \nnh spectra are also shown.
The map in the middle represents the $N$(H$_2$) distribution derived from
the $\it{Herschel}$ data in units of cm$^{-2}$ \cite[e.g.,][]{Andre2010}.
The lowest contour and the contour interval of the $N$(H$_2$) map are $3\times10^{22}$ cm$^{-2}$.
\label{fig:spectra}}
\end{figure*}

\begin{figure*}
\begin{center}
\includegraphics[scale=.4]{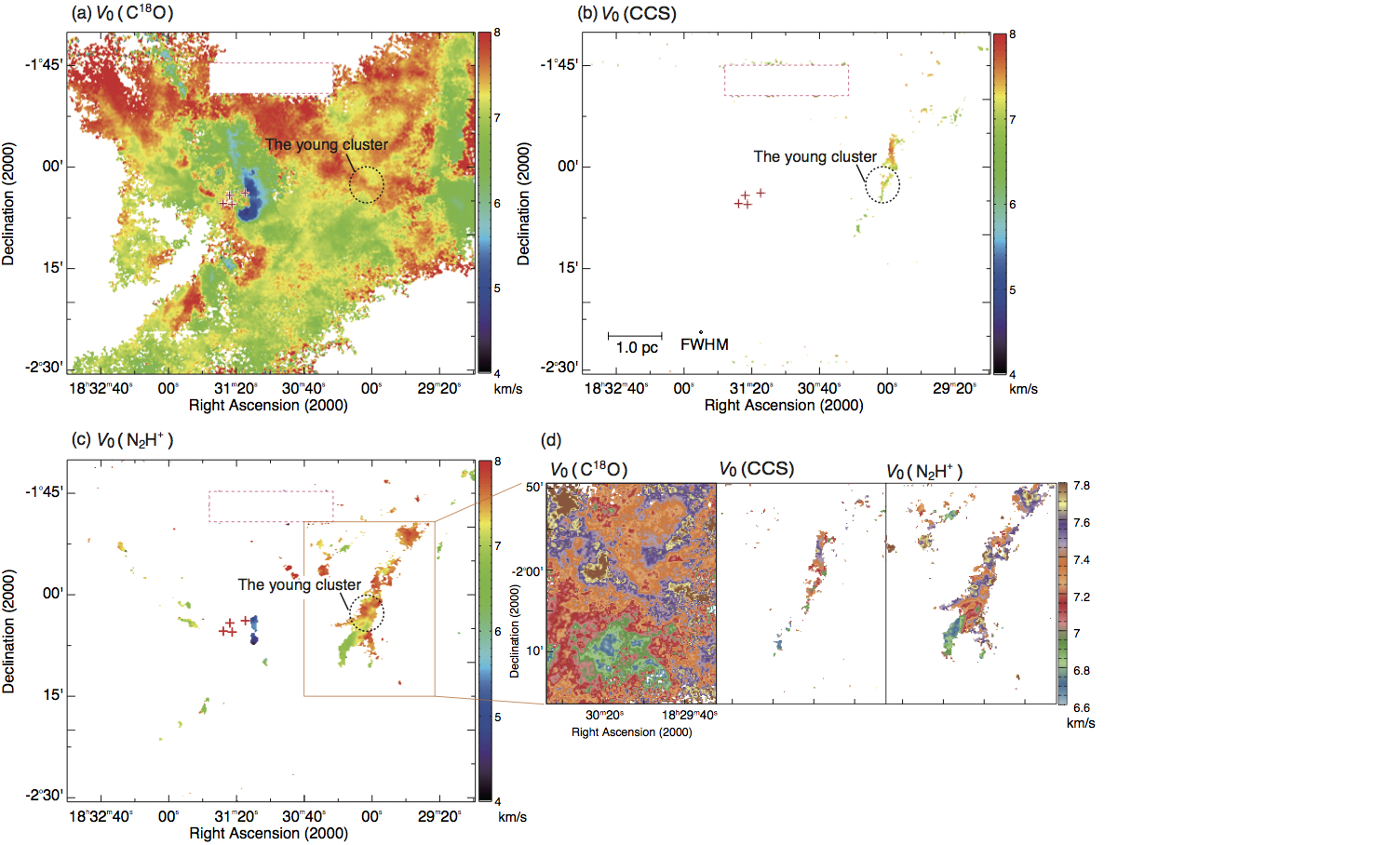}
\end{center}
\caption{
Mean velocity maps of the (a) C$^{18}$O, (b) CCS, and (c) N$_2$H$^{+}$ emission lines. 
The map of N$_2$H$^{+}$ was obtained by fitting all of the hyperfine lines (see text). 
(d) Close-up view of each map for the Serpens South region.
\label{fig:V0map}}
\end{figure*}

\begin{figure*}
\begin{center}
\includegraphics[scale=.4]{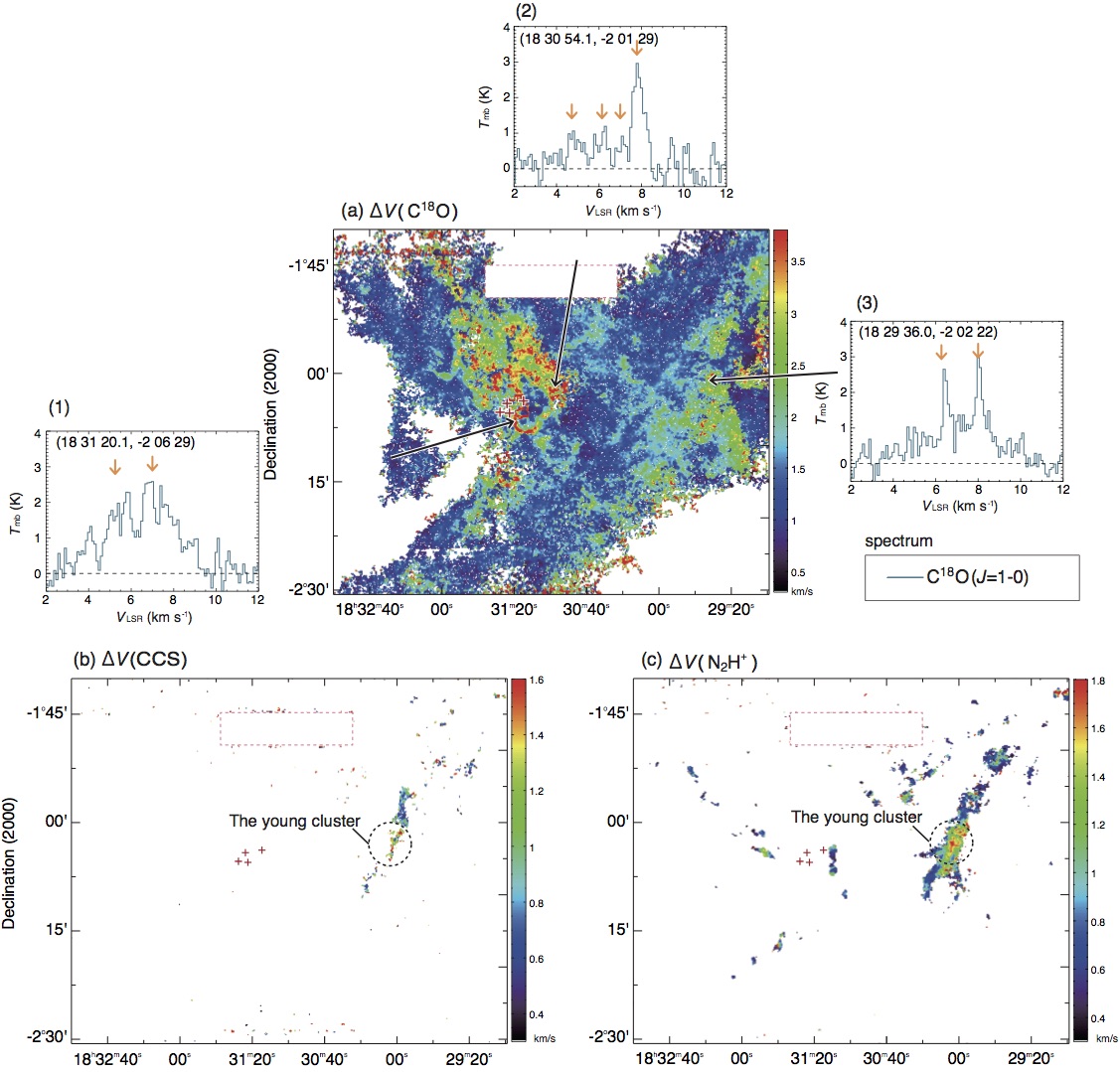}
\end{center}
\caption{
Velocity width (FWHM) maps of the (a) C$^{18}$O, (b) CCS, and (c) N$_2$H$^{+}$ emission lines.
In panel (a), the C$^{18}$O spectra observed at some positions indicated by the black arrows are shown.
Orange arrows denote distinct velocity components seen in the spectra.
\label{fig:dV}}
\end{figure*}

\begin{figure*}
\begin{center}
\includegraphics[scale=0.4]{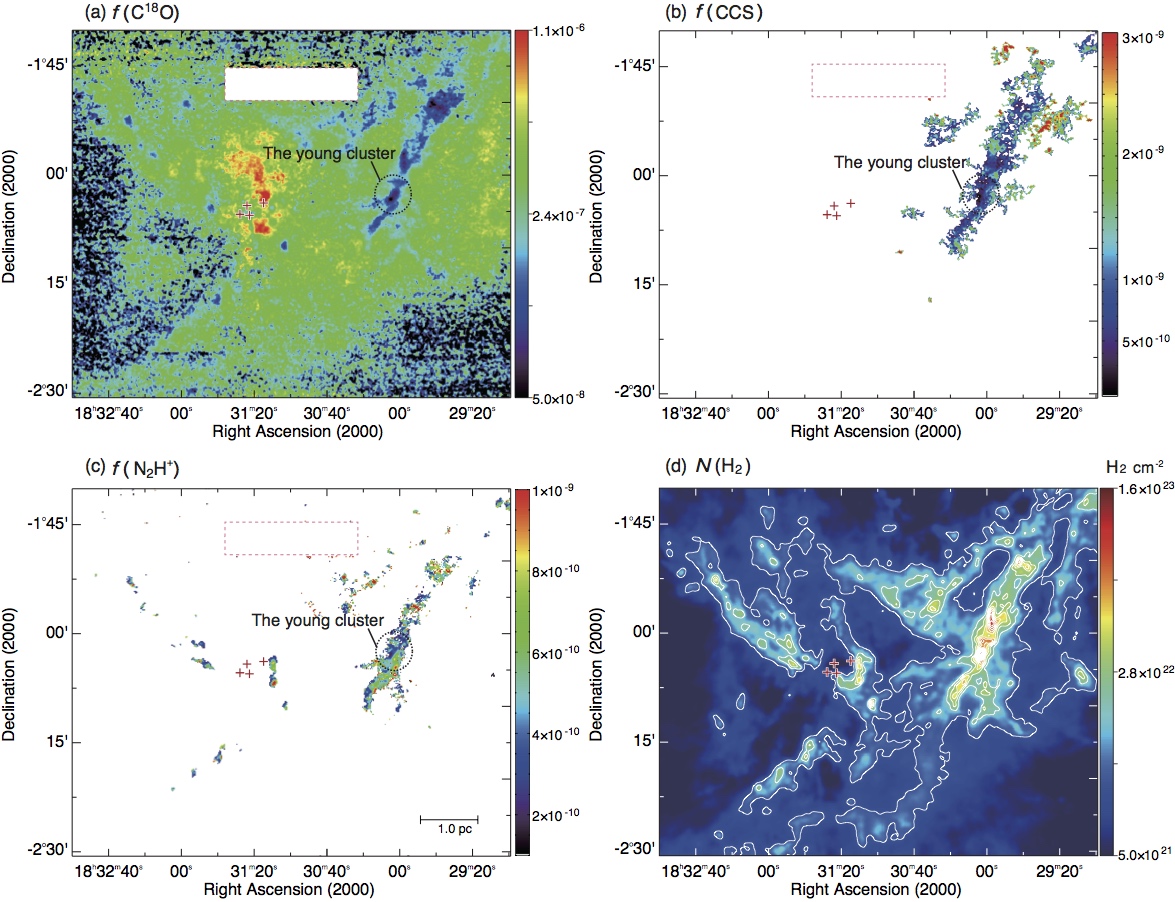}
\end{center}
\caption{
Fractional abundance maps of (a) C$^{18}$O, (b) CCS, and (c) N$_2$H$^{+}$.
We also show the $N$(H$_2$) map \cite[e.g.,][]{Andre2010} in panel (d).
The lowest contour and contour interval for the $N$(H$_2$) map are $1\times10^{22}$ cm$^{-2}$.
\label{fig:fmap}}
\end{figure*}

\begin{figure*}
\begin{center}
\includegraphics[scale=.4]{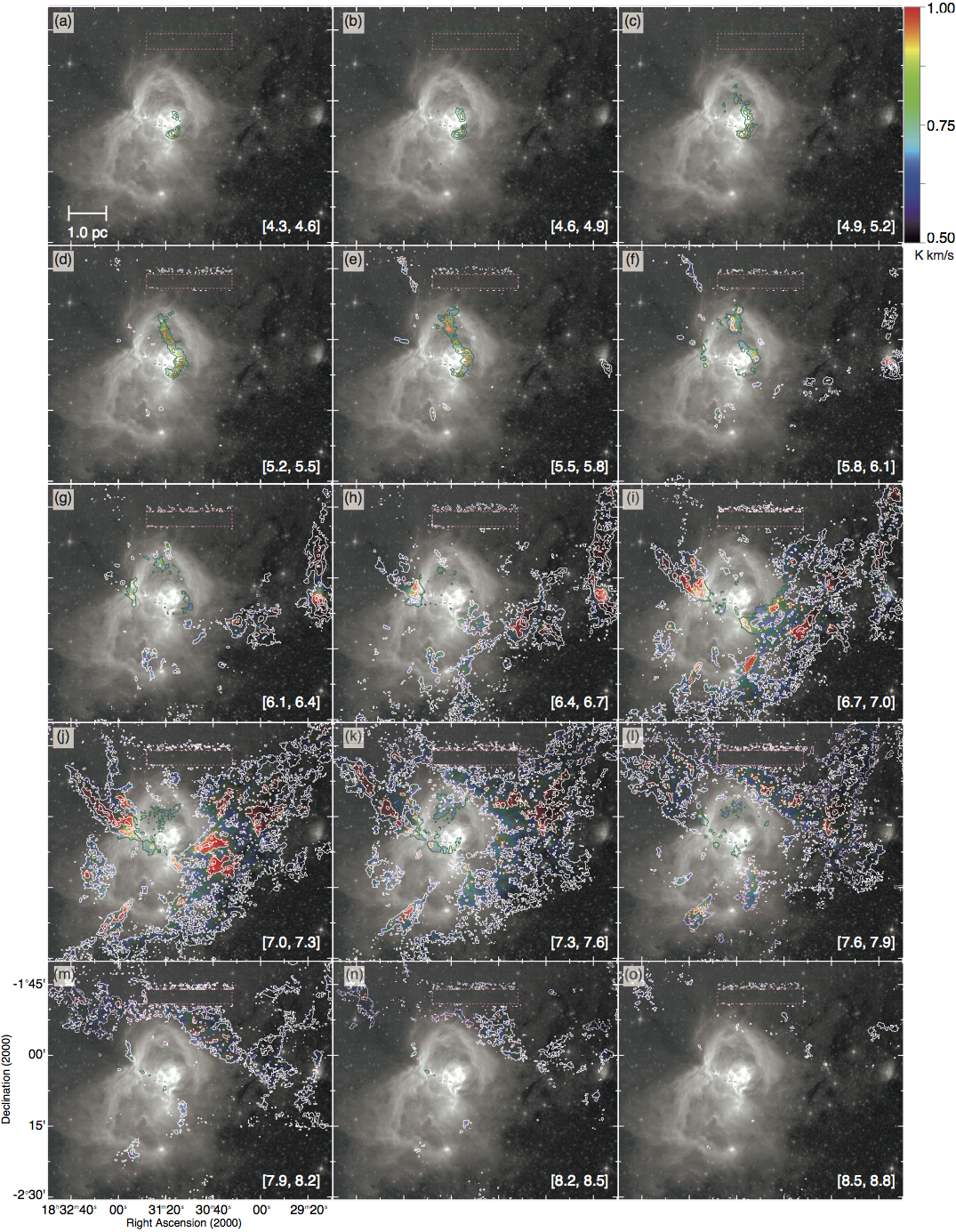}
\end{center}
\caption{
Channel maps of the C$^{18}$O emission made at every 0.3 km s$^{-1}$
in the velocity range from 4.3 km s$^{-1}$ to 8.8 km s$^{-1}$. Each panel is
overlaid on the image shown in figure \ref{fig:spitzer}. 
The velocity range (in units of km s$^{-1}$) used for the integration is shown in the brackets in each panel.
The lowest contour level and the contour interval are 0.5 K km s$^{-1}$.
The color scale for the C$^{18}$O intensity is indicated in the top right corner.
\label{fig:channel1}}
\end{figure*}

\begin{figure*}
\begin{center}
\includegraphics[scale=0.5]{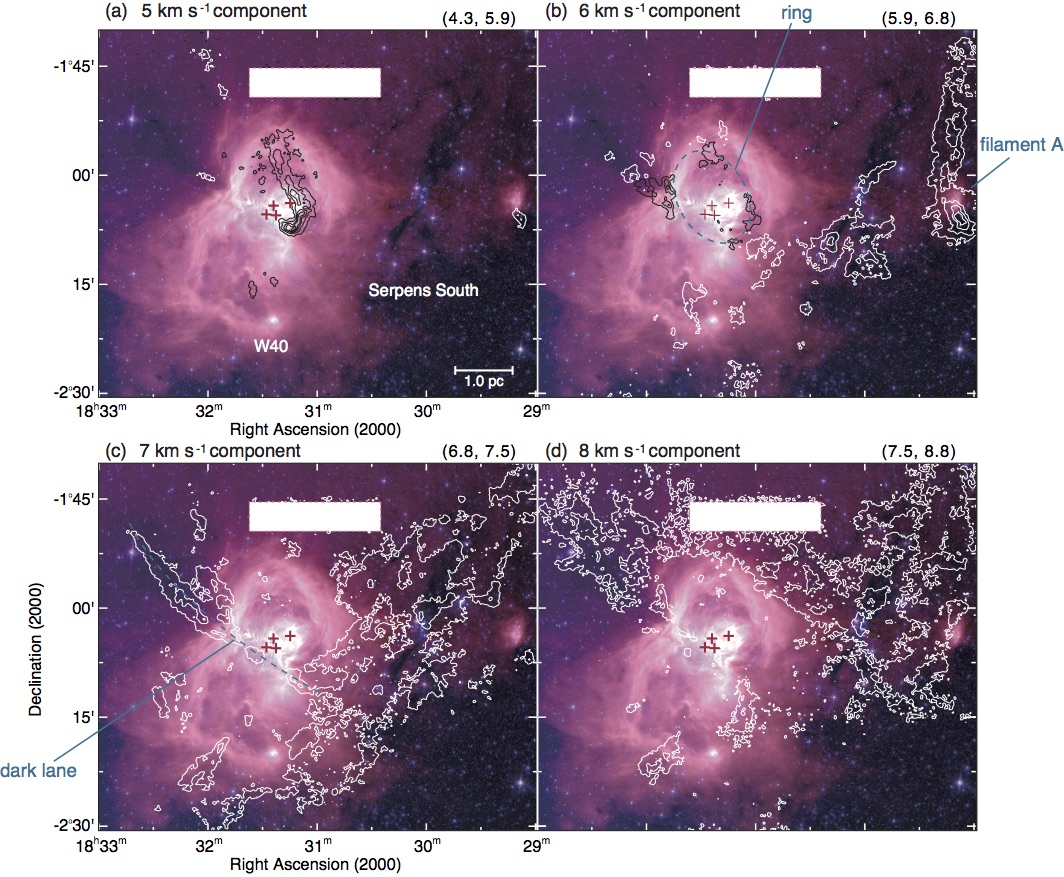}
\end{center}
\caption{
The \eco intensity distributions of the four velocity components shown by the black/white contours (see text).
The lowest contour and the contour interval are 1.4 K km s$^{-1}$.
The background image is the same as that in figure \ref{fig:spitzer}.
Velocity ranges in units of km s$^{-1}$ used for the integration of each component 
are shown in the parentheses above each panel.
\label{fig:4comp1}}
\end{figure*}

\begin{figure*}
\begin{center}
\includegraphics[scale=0.5]{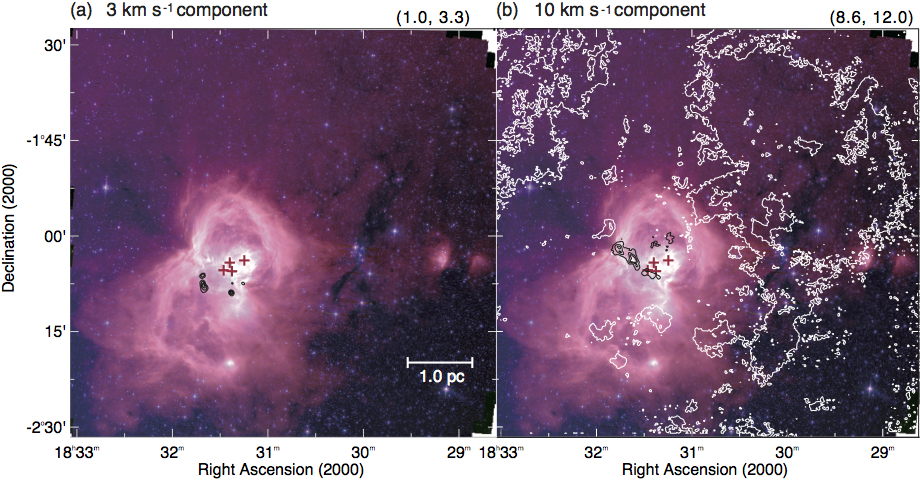}
\end{center}
\caption{
Distributions of the 3 km s$^{-1}$ component and the 10 km s$^{-1}$ component identified based on the $^{13}$CO data (see text). 
Velocity ranges in units of km s$^{-1}$ used for the integration of each component 
are shown in the parentheses above each panel.
The lowest contour and the contour interval are 3.0 K.
The background images are the same as that in figure \ref{fig:spitzer}.
\label{fig:2comp}}
\end{figure*}

\begin{figure*}
\begin{center}
\includegraphics[scale=0.4]{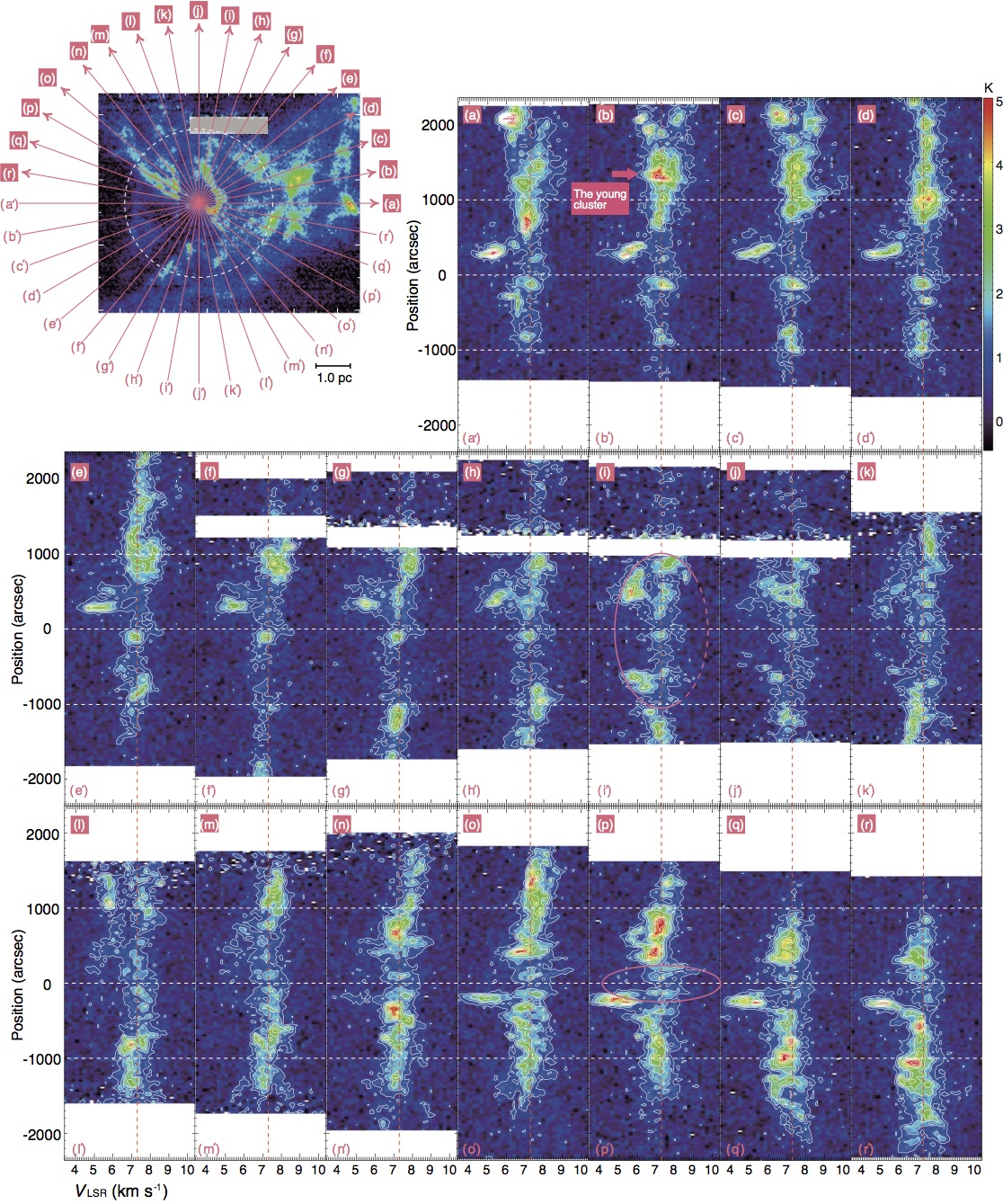}
\end{center}
\caption{
Position-velocity diagrams of the C$^{18}$O emission line taken along the cuts (a)--(r)
shown in the C$^{18}$O intensity map at the top-left panel.
The lowest contour and the contour interval are both 0.6 K.
The horizontal-vertical red broken line indicates the systemic velocity ($V_{\rm LSR}=7.3$ km s$^{-1}$).
The white broken lines indicate the position of IRS1A South as well as the positions
separated by $\pm1000\arcsec$ from the source roughly corresponding to the boundary
of the \Hii region.
Two elliptical structures are marked with a pink ellipse in panels i and p.
\label{fig:PV1}}
\end{figure*}

\begin{figure*}
\begin{center}
\includegraphics[scale=0.5]{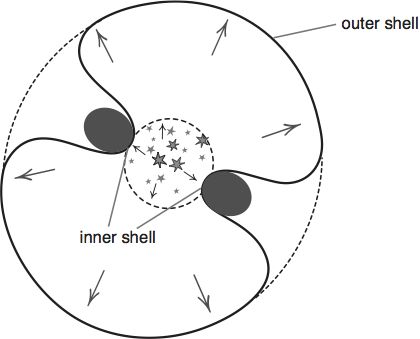}
\end{center}
\caption{
A schematic representation of the W40 \Hii region showing that the two shells were created by the expansion of the \Hii region.
The small inner shell is found in the vicinity of the \Hii region, and the other is the larger outer shell corresponding to the boundary of the \Hii region.
\label{fig:hourglass}}
\end{figure*}

\begin{figure*}
\begin{center}
\includegraphics[scale=0.4]{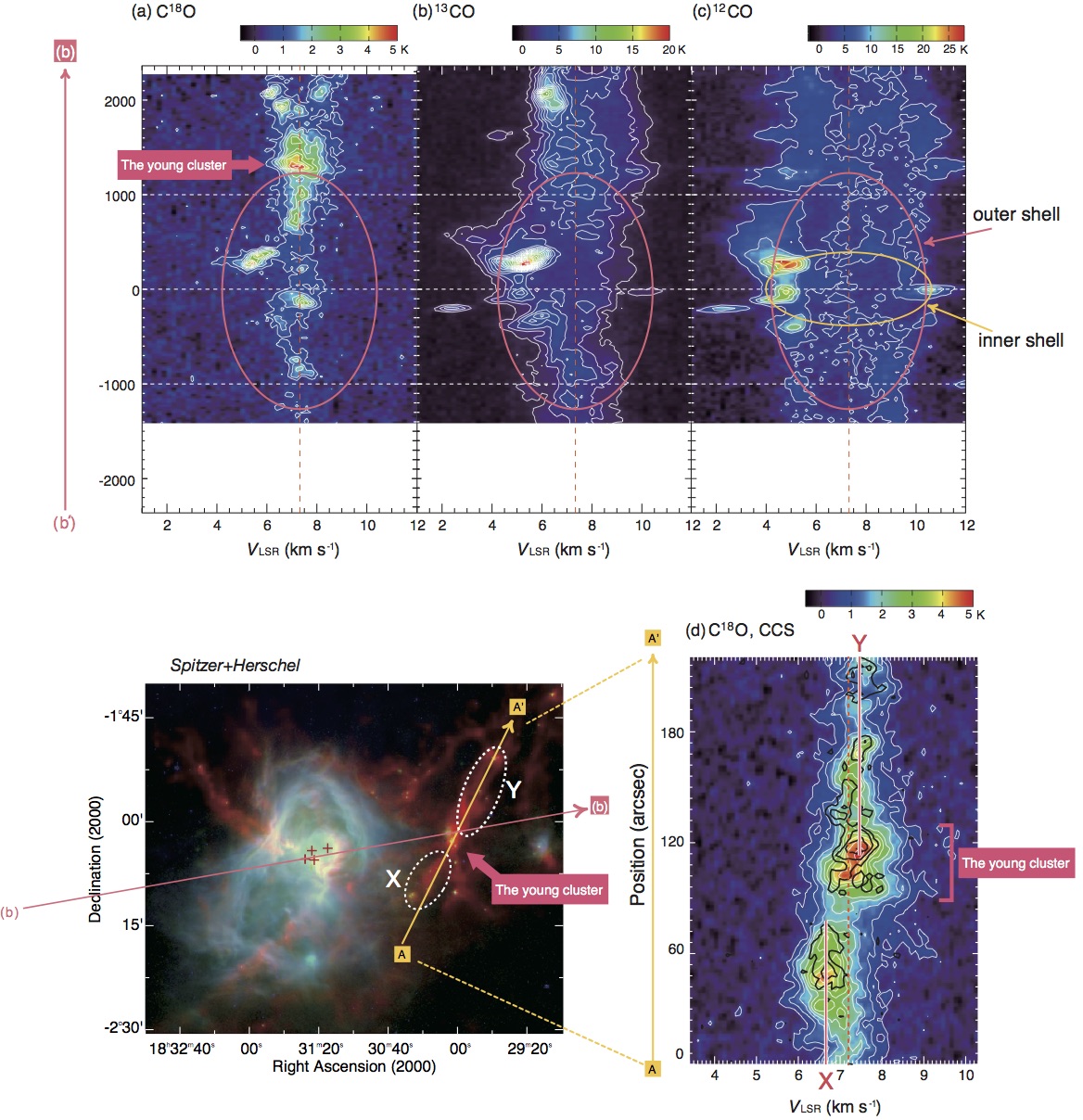}
\end{center}
\caption{
%
(a) PV diagram of the \eco emission line (the same diagram as figure \ref{fig:PV1}b).
PV diagram of (b) the $^{13}$CO emission line
and (c) the $^{12}$CO emission line taken along the same cut labeled b'-b.
The lowest contour and contour interval are 1.0 K km s$^{-1}$ for panel (b) 
and 4.0 K km s$^{-1}$ for panel (c).
(d) PV diagram of the C$^{18}$O emission line (color scale + white contour) and CCS emission line (black contour)
taken along the cut A-A'.
The lowest contour and contour interval are 0.6 K km s$^{-1}$ for \eco and 0.4 K for CCS.
The vertical broken line indicates the systemic velocity ($V_{\rm LSR}=7.3$ km s$^{-1}$) for panes a -- d.

\label{fig:PV2}}
\end{figure*}

\begin{figure*}
\begin{center}
\includegraphics[scale=0.4]{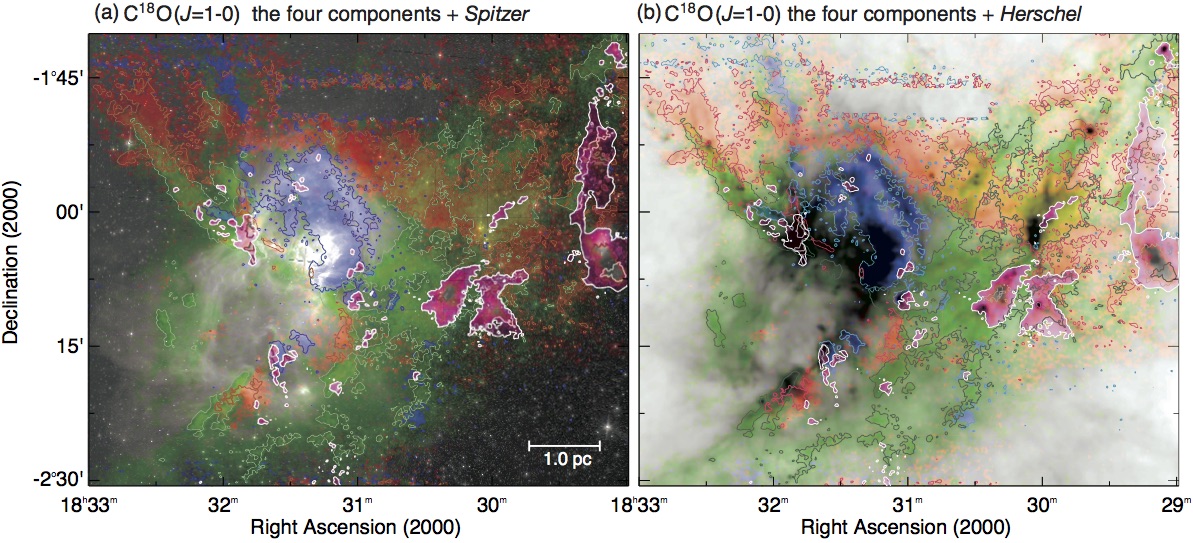}
\end{center}
\caption{
Composite color image of W40 and Serpens South made from the integrated intensity maps, with the 5 km s$^{-1}$ component in blue, the 6 km s$^{-1}$ component in pink, and the 7 km s$^{-1}$ component in green, and the 8 km s$^{-1}$ component in red, respectively. 
The background image is the same as that in figure \ref{fig:spitzer} for panel (a), 
and the $\it{Herschel}$ 250$\micron$ image for panel (b).
\label{fig:4comp2}}
\end{figure*}

\begin{figure*}
\begin{center}
\includegraphics[scale=0.2]{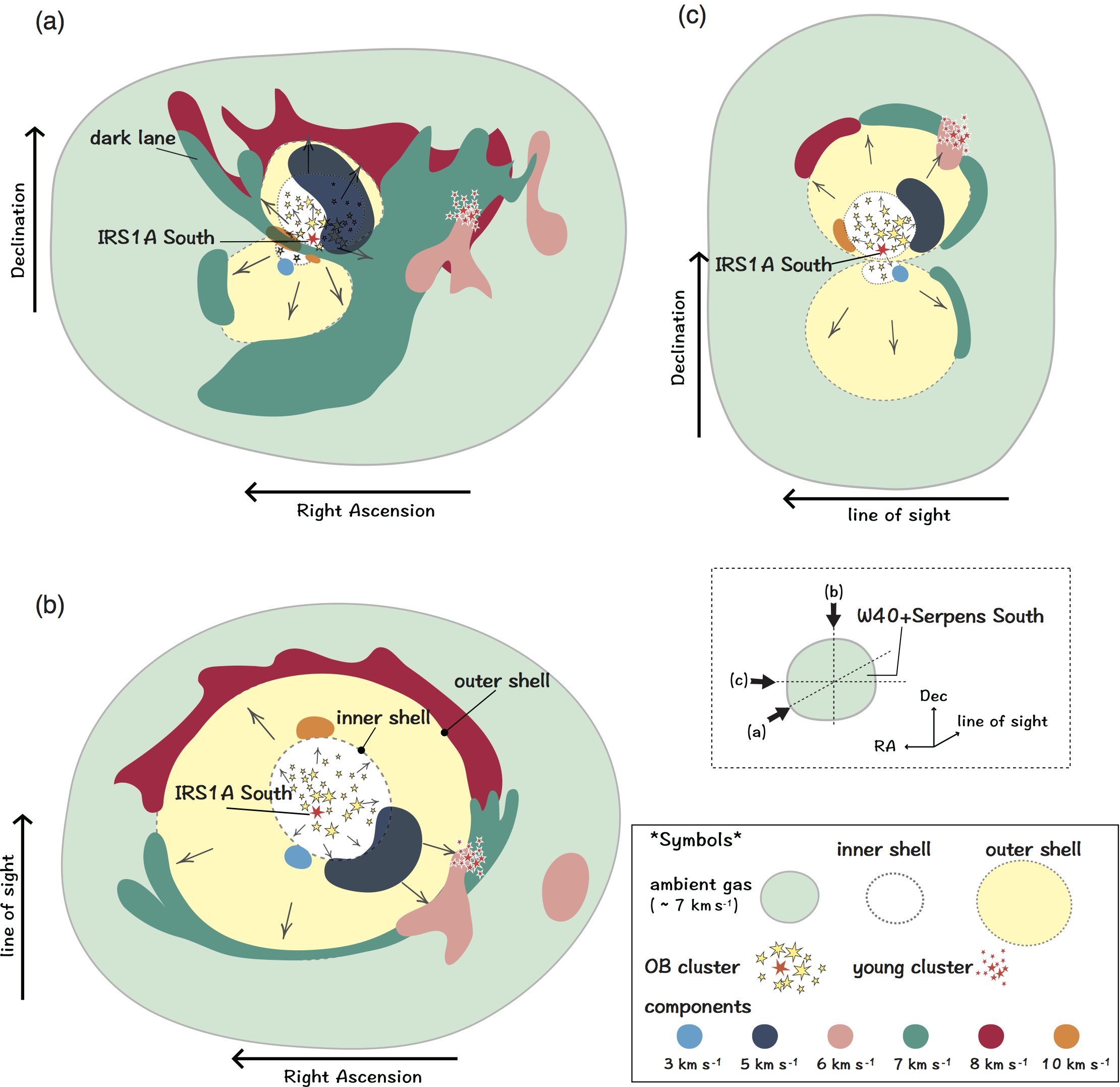}
\end{center}
\caption{
Schematic illustration of the proposed model for the three-dimensional geometry of the W40 and Serpens South complex.
As indicated in the box drawn by the broken line, panel (a) shows the view as is observed on the sky (e.g., see figure \ref{fig:4comp2}), panel (b) shows the view observed from the declination axis orthogonal to the line of sight, and panel (c) shows the view observed from the right ascension axis.
There are two expanding shells created by the \Hii region (see figure\ref{fig:hourglass}).
In the model, the 3 and 5 km s$^{-1}$ components are located on the near side of the inner shell, and the 10 km s$^{-1}$ component is located on the far side of the inner shell. 
The 6 and 7 km s$^{-1}$ component are located around the surface of the outer shell.
Both of the components associated with the young cluster of Serpens South are likely to be interacting with the outer shell.
The 8 km s$^{-1}$ component is located on the far side of the outer shell.
\label{fig:model}}
\end{figure*}


\end{document}